\documentclass[fleqn,usenatbib]{mnras}

\usepackage{newtxtext,newtxmath}
\usepackage{comment}

\usepackage[T1]{fontenc}
\usepackage{ae,aecompl}

\usepackage{graphicx}	
\usepackage{amsmath}	
\usepackage{amssymb}	

\title[TNG Morphologies]{The Hubble Sequence at $z\sim0$ in the IllustrisTNG simulation with deep learning}

\author[Huertas-Company et al.]{
Marc Huertas-Company,$^{1,2}$\thanks{E-mail: marc.huertas@obspm.fr}
 Vicente Rodriguez-Gomez$^{3}$, Dylan Nelson$^{4}$, 
\newauthor Annalisa Pillepich$^{5}$, Mariangela Bernardi$^{6}$, Helena Dom\'inguez-S\'anchez$^{6}$, Shy Genel$^{7}$
\newauthor Ruediger Pakmor$^{8}$, Gregory F. Snyder$^{9}$, Mark Vogelsberger$^{10}$
\\
$^{1}$ Instituto de Astrof\'isica de Canarias (IAC); Departamento de Astro\'isica, Universidad de La Laguna (ULL), E-38200, La Laguna, Spain\\
$^{2}$ LERMA, Observatoire de Paris, CNRS, PSL, Universit\'e Paris Diderot, France\\
$^{3}$Instituto de Radioastronom\'ia y Astrof\'isica, Universidad Nacional Aut\'onoma de M\'exico, Apdo. Postal 72-3, 58089 Morelia, Mexico\\
$^{4}$Max-Planck-Institut f{\"u}r Astrophysik, Karl-Schwarzschild-Str. 1, 85741 Garching, Germany\\
$^{5}$Max-Planck-Institut f{\"u}r Astronomie, K{\"o}nigstuhl 17, 69117 Heidelberg, Germany\\
$^{6}$Department of Physics and Astronomy, University of Pennsylvania, Philadelphia, PA 19104, USA\\
$^{7}$Center for Computational Astrophysics, Flatiron Institute, 162 Fifth Avenue, New York, NY 10010, USA\\
$^{8}$Heidelberg Institute for Theoretical Studies, Schloss-Wolfsbrunnenweg 35, 69118 Heidelberg, Germany\\
$^{9}$Space Telescope Science Institute, 3700 San Martin Dr, Baltimore, MD 21218, USA\\
$^{10}$Kavli Institute for Astrophysics and Space Research, Department of Physics, MIT, Cambridge, MA, 02139, USA
}

\date{Accepted XXX. Received YYY; in original form ZZZ}

\pubyear{2015}

\begin{document}
\label{firstpage}
\pagerange{\pageref{firstpage}--\pageref{lastpage}}
\maketitle

\begin{abstract}
We analyze the optical morphologies of galaxies in the IllustrisTNG simulation at $z\sim0$ with a Convolutional Neural Network trained on visual morphologies in the Sloan Digital Sky Survey. We generate mock SDSS images of a mass complete sample of $\sim12,000$ galaxies in the simulation using the radiative transfer code SKIRT and include PSF and noise to match the SDSS r-band properties. The images are then processed through the exact  same neural network used to estimate SDSS morphologies to classify simulated galaxies in four morphological classes (E, S0/a, Sab, Scd).  The CNN model finds that $\sim95\%$ of the simulated galaxies fall in one the four main classes with high confidence. The mass-size relations of the simulated galaxies divided by morphological type also reproduce well the slope and the normalization of observed relations which confirms the realism of optical morphologies in the TNG suite. However, the Stellar Mass Functions decomposed into different morphologies still show significant discrepancies with observations both at the low and high mass end. We find  that the high mass end of the SMF is dominated in TNG by massive disk galaxies while early-type galaxies dominate in the observations according to the CNN classifications. The present work highlights the importance of detailed comparisons between observations and simulations in comparable conditions.   
\end{abstract}




\section{Introduction}
Understanding the physical processes that lead to the diversity of galaxy morphologies we see in today's Universe, i.e. the Hubble Sequence, is still a major goal in the field of galaxy evolution. Until recently, numerical simulations struggled to simulate galaxies with realistic morphologies. An improvement of spatial resolution together with more accurate numerical codes and treatments of physical processes, have triggered the emergence of hydrodynamical cosmological simulations which produce galaxies with a variety of morphologies in the local universe (e.g. \citealp{2015MNRAS.446..521S, 2015ApJ...804L..40G, 2015MNRAS.452.1502D}). This allows us to move from a qualitative to a more quantitative approach in which the number densities as well as other scaling relations of different simulated morphologies can be compared to observations. It requires to consistently measure morphologies in the simulations as done in the observations. Radiative transfer codes enable one to forward model the simulation outputs and produce mock observations under some assumptions on the conversion from mass to light as well as on the absorption of light by dust.

However, there have been few works that precisely quantify the detailed morphologies of the simulated galaxies.  This is partly due to the fact that quantifying galaxy morphology for large numbers of galaxies has also remained an elusive problem in the observations. An interesting attempt to process simulated galaxies with the same methodology was carried out by \cite{2018ApJ...853..194D} in the framework of the Galaxy Zoo project \citep{2011MNRAS.410..166L}. They classified a complete sample of simulated galaxies from the Illustris simulation using the citizen science approach developed by the Galaxy Zoo collaboration and showed that simulated galaxies still presented important differences with respect to observed ones They found in particular that simulated galaxies present more substructures than observed ones, especially at lower masses. One caveat of this approach is that it is very  time consuming to consistently process different simulations with the same methodology. Other works have then  followed a more automated approach.  \cite{2017MNRAS.467.1033B, 2017MNRAS.467.2879B} performed bulge-disc decompositions of mock Illustris galaxies finding also significant differences with the observations. They find in particular a deficit of bulge-dominated galaxies at the high mass end which provokes that the size-luminosity relations of Illustris galaxies present higher normalizations and smaller slopes than for real galaxies. More recently, \cite{2019MNRAS.483.4140R} used parametric and non-parametric morphological proxies of the new IllustrisTNG galaxies (see also \citealp{2015MNRAS.454.1886S} for a similar approach). They measured a significant improvement in terms of scaling relations compared to the original run but some discrepancies remain on the slopes and normalizations of the mass-size relations of early and late-type galaxies. Galaxies from the EAGLE simulation have also recently been processed through radiative transfer codes~\citep{2017MNRAS.470..771T} and used for example to identify barred systems~\citep{2018MNRAS.481.2951E}. However, there is no precise quantification of the morphological mix. These approaches have revealed interesting trends, while at the same time motivating more sophisticated comparison techniques.

In recent years, machine learning and more precisely deep learning, has emerged as an extremely efficient tool to estimate detailed visual like morphologies from images (e.g. \citealp{2015MNRAS.450.1441D,2015ApJS..221....8H,2018MNRAS.476.3661D}). It offers an interesting and fast approach to consistently compare theory and observations since it becomes possible to efficiently apply the exact same methods to simulations and observations and obtain accurate detailed morphologies. The realism of the simulated galaxies can be quantified in detail. 

The main purpose of this work is thus to quantify the detailed optical \emph{visual} morphologies of the new TNG100 simulation of the IllustrisTNG suite~\citep{2018arXiv181205609N} and compare with observations. We use to that purpose a Convolutional Neural Network trained on the SDSS \citep{2018MNRAS.476.3661D} to classify a mass selected ($Log(M_*/M_\odot)>9.5$) sample of $\sim12,000$ simulated galaxies at $z=0.05$ in four major morphological types (E, S0, Sab, Scd) in the SDSS r-band. Bayesian Neural Networks are used to quantify the similarities between observations and simulations.  We then analyze the Stellar Mass Functions and mass-size relations of simulated galaxies divided by morphological type and compare with observations.

The paper proceeds as follows. Sections~\ref{sec:sims} and~\ref{sec:obs} describe the simulated and observational samples used in this work. We then describe in section~\ref{sec:deep} the main methodology used to quantify galaxy morphologies with CNNs. In section~\ref{sec:bnns} we discuss the similarity between simulated and observed morphologies from the machine learning perspective. The main results regarding the scaling relations and stellar mass functions of simulated and observed galaxies are presented in sections~\ref{sec:scaling} and~\ref{sec:SMFs} respectively.
\section{Simulations}
\label{sec:sims}
\subsection{The IllustrisTNG simulation}

The IllustrisTNG Project \citep{2018arXiv181205609N, 2018MNRAS.480.5113M, 2018MNRAS.477.1206N, 2018MNRAS.475..676S, 2018MNRAS.473.4077P} is a suite of magneto-hydrodynamic cosmological simulations performed with the moving-mesh code AREPO~\citep{2010ARA&A..48..391S, 2011MNRAS.418.1392P, 2016MNRAS.455.1134P}. See \cite{2017MNRAS.465.3291W, 2018MNRAS.473.4077P} for a description of the TNG simulation model which is an improved version of the original Illustris simulation~\citep{2014Natur.509..177V, 2014MNRAS.444.1518V, 2014MNRAS.445..175G, 2015MNRAS.452..575S}. The IllustrisTNG model was especially designed to match some key observables: (i) the global star formation rate density at $z=0-8$, (ii) the galaxy mass function at $z=0$, (iii) the stellar-to-halo mass relation at $z=0$, (iv) the black hole-to-stellar mass relation at $z=0$, (v) the halo gas fraction at $z=0$, and (vi) galaxy sizes at $z=0$. In this work, we use the highest resolution version of TNG100, which follows the evolution of $2 \times 1820^3$ resolution elements within a periodic cube measuring $75h^{-1}\simeq110.7$ Mpc. The main differences with respect to the first Illustris run consist of a new active galactic nucleus (AGN) feedback model that operates at low accretion rates \citep{2017MNRAS.465.3291W} and a reworking of the galactic winds~\citep{2018MNRAS.473.4077P}, and the inclusion of magnetic fields~\citep{2011MNRAS.418.1392P}. This simulation has been recently made publicly available~\citep{2018arXiv181205609N}.

\subsection{Dataset and Synthetic images}

In this work, we consider a single simulation snapshot at $z = 0.0485$ (snapshot 95) as done in~\cite{2019MNRAS.483.4140R}. The redshift is consistent with the average redshift of the observations comparison sample ($z_{med}=0.06$). Within this snapshot, we consider all simulated galaxies with $Log(M_*/M_\odot) > 9.5$ which is also roughly consistent with the stellar masses in the observational dataset detailed in the following section. The sample selected contains $12,468$ galaxies (hereafter TNG sample). We do not require here a perfect stellar mass and redshift match between observations and simulations since all properties will be explored at fixed stellar mass. It is important though that all types of galaxies are well represented in the training set used to train the CNNs. We will further explore this in section~\ref{sec:bnns}. \\

From the parent sample we create synthetic images for all the galaxies in the snapshot using the radiative code SKIRT~\citep{2011ApJS..196...22B}\footnote{http://www.skirt.ugent.be/root/index.html}. We refer the reader to~\cite{2019MNRAS.483.4140R} for full details on how the images are created. In short, each galaxy is observed from a unique random viewing angle perpendicular to the xy-plane of the simulation volume. The field of view of each image is equal to 15 times the (3D) stellar half-mass radius of the corresponding galaxy. The number of pixels is tuned to match the SDSS pixel scale ($0.396$ arcsec, $0.38$ kpc at $z=0.05$). The stellar populations are modeled with the \cite{2003MNRAS.344.1000B} stellar population models for stars older than 10Myrs. Younger stars are considered starbursting regions and are modeled with the MAPPINGS-III photoionization code~\citep{2008ApJS..176..438G}. All details of how parameters are set can be found in~\cite{2019MNRAS.483.4140R}. For computational reasons, dust is taken into account only if the fraction of star forming gas is above $1\%$ of the total baryonic mass. It is assumed for these objects that  the dust content is traced by the star-forming gas. A constant dust-to-metal ratio of 0.3 is also assumed. The final output is a 3D data cube for each galaxy, consisting of a full rest-frame SED for each pixel. We
then assume that the source is located at z = 0.0485 and generate the data cube that would be measured by a local observer, taking cosmological effects such as surface brightness dimming into account. Each SED is then multiplied by each of the SDSS filter curves (g,r,i,z) and integrated over the full wavelength range. We use only the r-band in this work. Instrumental effects are added by convolving the images with an SDSS PSF computed by averaging $\sim 1000$ stars\footnote{Trujillo, Infante private communication}.  Very bright ($<$7 mag) stars (although saturated and showing bleeding patterns in the central part) were used for creating the outermost regions. Stacking of stars with intermediate brightness ($\sim$9 mag) were used for building the intermediate region of the PSF whereas faint ($>$14 mag; non-saturated stars) were used for the central core. Given that the galaxies considered in this work are nearby, well-resolved objects, the impact of eventual spatial variations of the PSF is negligible and are not taken into account. Realistic noise distributions taken from empty regions from the SDSS DR7 are also added to the images to match the S/N distribution of the observed sample. Note that in the procedure to generate images, neighboring galaxies are not included. The images only contain the central galaxy. This is a difference with respect to the observational sample described in the following section. However, since the training of the network model is performed in the observations as described in section~\ref{sec:deep}, this should not represent a problem for the reliability of the classifications obtained in the simulations. In any case, in future work, we plan to include companion galaxies in the simulated stamps to confirm that no bias is introduced.
\section{Observations}
\label{sec:obs}

\subsection{SDSS parent sample}

The observational sample used in this work comes from the $670,722$ galaxies selected by \cite{2015MNRAS.446.3943M} (hereafter the M15 sample) from the SDSS DR7 spectroscopic sample. We refer the reader to the aforementioned work for all the details regarding the selection. Very briefly, galaxies are selected from the SDSS DR7 database according to three main criteria: (1) the extinction corrected r band Petrosian magnitude is between 14 and 17.77. The limit at the bright end is to avoid large nearby galaxies which are typically split into multiple objects in the SDSS catalogue. The faint end limit is the lower limit for completeness of the SDSS spectroscopic survey \citep{2002AJ....124.1810S}; (2) the photometric pipeline classified the object as a galaxy; and (3) the spectrum was also identified as a galaxy. Some further cleaning of very nearby objects ($z<0.0005$) and objects with catastrophic photometric redshifts results in sample of $670,722$ galaxies. The median redshift of this sample is $\sim0.09$ and goes up to $z\sim0.25$ which is slightly higher that the one of the IllustrisTNG snapshot described in the previous section. In order to reduce the impact of possible morphological evolution, we select for this work only objects with $z<0.1$. This additional selection results in a final sample of $328,709$ galaxies. The volume probed by the observational sample is roughly 40 times larger than the simulated TNG volume. The morphological mix, which is the main property we aim to measure in this work, might eventually change in small volumes, especially at the high mass end. This is addressed in detail in section~\ref{sec:SMFs}.

\subsection{Structural parameters and stellar masses}
A large number of derived quantities exist for the dataset described above. In particular we use for this work the Stellar Masses computed for all galaxies~\citep{2017MNRAS.467.2217B}\footnote{catalog available at: \url{http://alan-meert-website-aws.s3-website-us-east-1.amazonaws.com/fit_catalog/download/index.html}}. Stellar masses are derived using a \cite{2003PASP..115..763C} IMF and the M/L ratio from \cite{2015ApJ...804L...4M}. We do not include any variation in the IMF. We refer to~\cite{2018MNRAS.475..757B} for the implications of IMF gradients in the mass. The luminosity comes from the Sersic best fit models derived for all galaxies in the catalog (see \citealp{2015MNRAS.446.3943M} for more details). We also use the effective radii estimated through fitting Sersic models when comparing the scaling relations of observed and simulated galaxies. \cite{2018MNRAS.476.3661D} also derived detailed visual morphologies for this sample with CNNs. However, in this work we perform a new training to ensure that the neural networks are trained on data with similar properties to the simulations. 

\section{Deep learning r band \emph{visual} morphologies}
\label{sec:deep}

We train a CNN on the visual morphologically classified sample of \cite{2010ApJS..186..427N} (hereafter N10 sample). The catalog contains detailed morphologies of $\sim 14,000$ galaxies performed by two professional astronomers. The authors associate to every galaxy in the sample a numeric value (T-Type) indicating the morphological type spanning from $-5$ (Elliptical) to $10$ (Irr) - see Table 3 in \cite{2010ApJS..186..427N}. We used this dataset instead of the Galaxy Zoo catalog even if it is smaller in size because the classification reflects well the standard Hubble Sequence which is not the case in the Galaxy Zoo classification tree. 

We use the same vanilla architecture as in~\cite{2018MNRAS.476.3661D}. However we performed a new training instead of directly using the published catalog because \cite{2018MNRAS.476.3661D} used JPEG images from the SDSS to train the networks. Since we want to make sure in this work that the neural networks see exactly the same data in the observations and in the simulations, we use fits r-band images as input for the training. The input stamps are of fixed size $128\times128$ ($\sim50\times 50$ kpc at $z=0.05$) which is larger than 2 effective radii of $\sim90\%$ of galaxies in the sample. Before being fed into the CNN, images are background subtracted by removing the median of the pixels in an empty region and normalized to the maximum value so that all images span a similar range between 0 and 1. We also tried other non linear normalizations such as hyperbolic sine to boost the signal in the outskirts of the galaxies. However this had no significant impact in the performance so we decided to keep a simpler linear normalization.   

The training strategy is also a bit different than in \cite{2018MNRAS.476.3661D}. In that work, they performed a regression on the T-type of the galaxy. We simplified the problem here into a hierarchical binary classification problem as done in~\cite{2011A&A...525A.157H} which is easier to evaluate and train (see \citealp{2019_Silla} for a review of hierarchical classifications). This is enough for our purposes since we will use only four main classes. We thus train 3 different binary classifiers with the same architecture. The first model (Model-1) is trained to separate early-type ($Ttype\leq0$) from late-type galaxies ($Ttype\geq1$). Model-1 delivers therefore a probability for a galaxy to be late-type: $P(Late)=1-P(Early)$.  A second model (Model-2) distinguishes Ellipticals ($Ttype\leq-3$) from S0/a's ($-3<Ttype<1$). Model-2 measures then the probability of being S0 with the prior that the galaxy is early-type: $P(S0/Early)=1-P(E/Early)$. Finally a third model (Model-3) splits objects between Sab, Sb galaxies ($1\leq Ttype<4$) and late-type spirals and irregulars ($Ttype\geq4$) with a training set made only of late-type galaxies. Model-3 estimates the probability for a galaxy to be a late-type spiral given that it is a late-type galaxy: $P(Scd/Late)=1-P(Sab/Late)$. The sizes of the training set decreases as one goes deeper into the tree, but it is enough to avoid over-fitting. 

\begin{figure}
\includegraphics[width=0.45\textwidth]{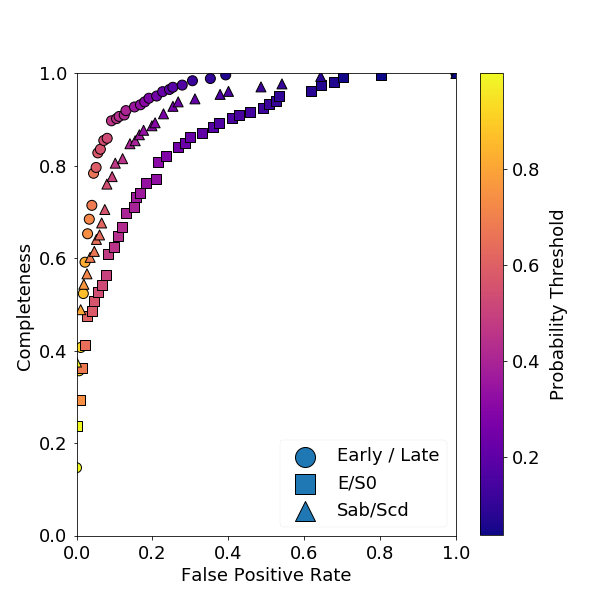}
\includegraphics[width=0.45\textwidth]{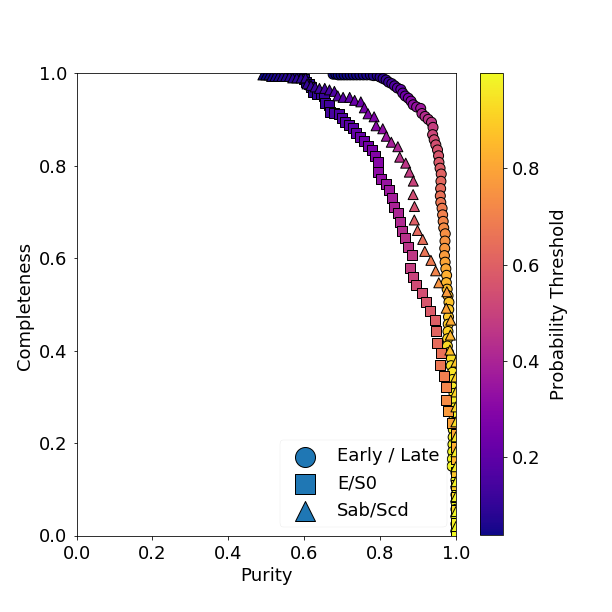}
\caption{Performance of the three morphological classifiers used in this work (see text for details). The top panel shows the ROC curve and the bottom panel the Precision - Recall curve (purity - completeness). The circle, square and triangle symbols indicate respectively the early/late, Sab/Scd and E/S0 classifiers.  The color bar indicates the corresponding probability threshold. See text for details.}
\label{fig:roc_curve}
\end{figure}

Figure~\ref{fig:roc_curve} shows the ROC (Receiver operating characteristic) and Precision-Recall (or Purity-Completeness) curves for the three classifications computed on a test set never used for training. In a binary classification, the ROC curve shows the fraction of false positives (i.e. in our case the fraction of galaxies classified as late-type among all galaxies with an early-type label) vs. the fraction of true positives (i.e. in our case, the fraction of galaxies classified as late-type among all galaxies with a late-type label). The true positive rate is typically called in astronomy completeness. Since the network outputs a probability and not a binary number, one can change the threshold to define positives (i.e. late-type galaxies). The smaller the threshold the larger the fraction of true positives but also the larger the fraction of false positives. This is shown in the ROC curve in which every point shows the fractions of false positives and true positives for different thresholds. The closer the curve gets to the top left corner, the more accurate the classifier is. For comparison, a random classifier will always have an equal fraction of false and true positives. The Precision-Recall curve is another indicator in which instead of plotting the fraction of false positives, we plot the fraction of true positives among all positive examples classified by the network (i.e. fraction with a true late-type label among all objects classified as late-type). The latter is a proxy for purity. For the P-R curve, both quantities need therefore to be as close as possible to one. 

 As expected the best accuracy is achieved for the first classifier (early vs. late) with a $\sim90\%$ purity and completeness consistent with previous works. The accuracy slightly decreases when more detailed morphologies are considered but still remains above $80\%$ in both purity and completeness. This is achieved for a typical probability threshold around 0.5 as expected for a well calibrated classifier. We emphasize that the main purpose of this work is not to obtain the best possible match with a human based classification but to apply exactly the same model to observations and simulations. 

We then use the three models to classify both the TNG and M15 samples. Every galaxy in both samples has therefore 3 different probabilities. The output of Model-2 has however little meaning for galaxies classified as late-type by Model-1. The same is true for the output of Model-3 and galaxies classified as early-type by Model-1. Therefore we associate four probabilities to every galaxy using the Bayes theorem:
\begin{equation}
\begin{split}
P(E)=P(Early)\times P(E/Early)  \\
P(S0)=P(Early)\times P(S0/Early) \\
P(Sab)=P(Late)\times P(Sab/Late) \\
P(Scd)=P(Late)\times P(Scd/Late)
\end{split}
\end{equation}

\begin{figure*}
\begin{tabular}{|c|c|}
      \hline
      \includegraphics[width=0.45\textwidth]{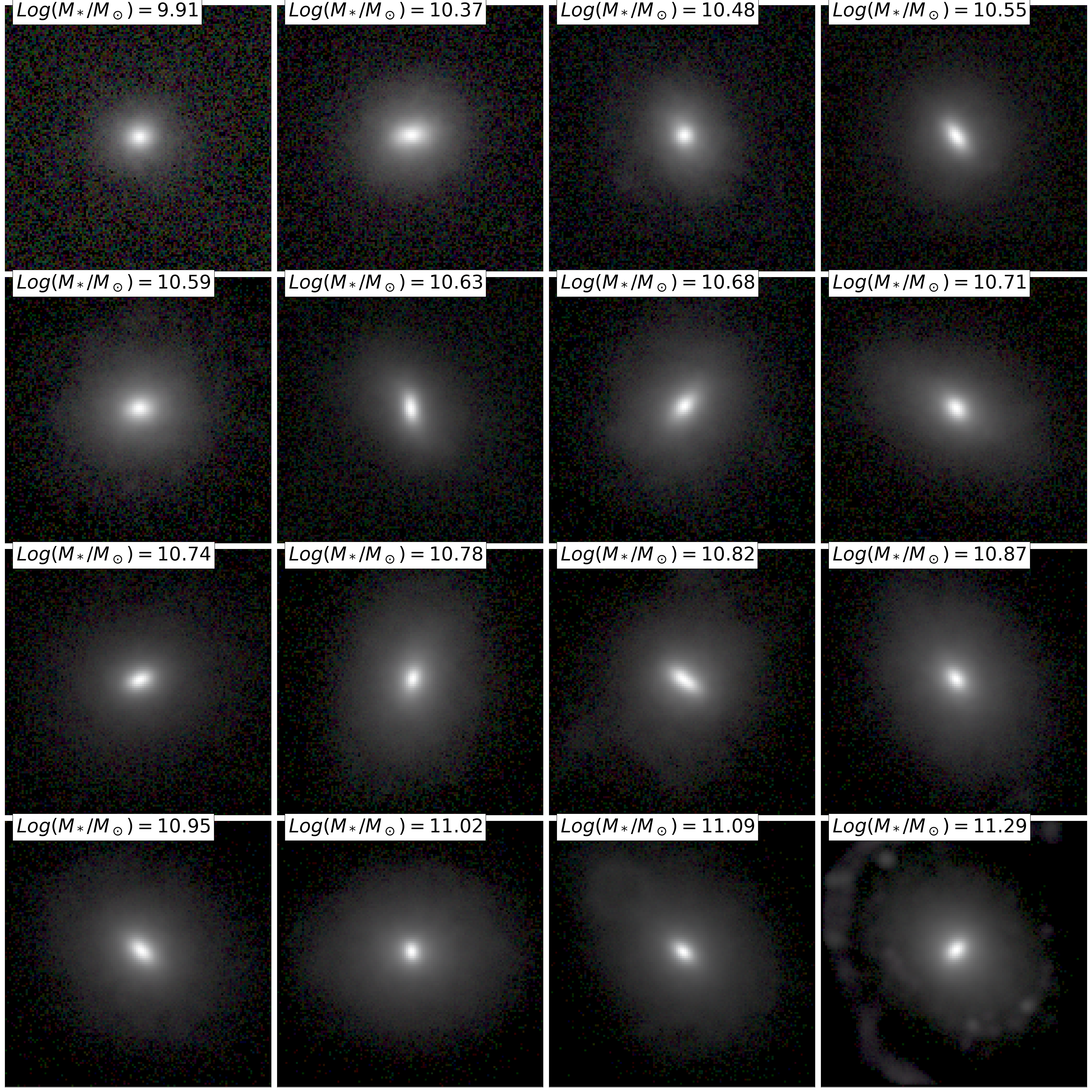} & \includegraphics[width=0.45\textwidth]{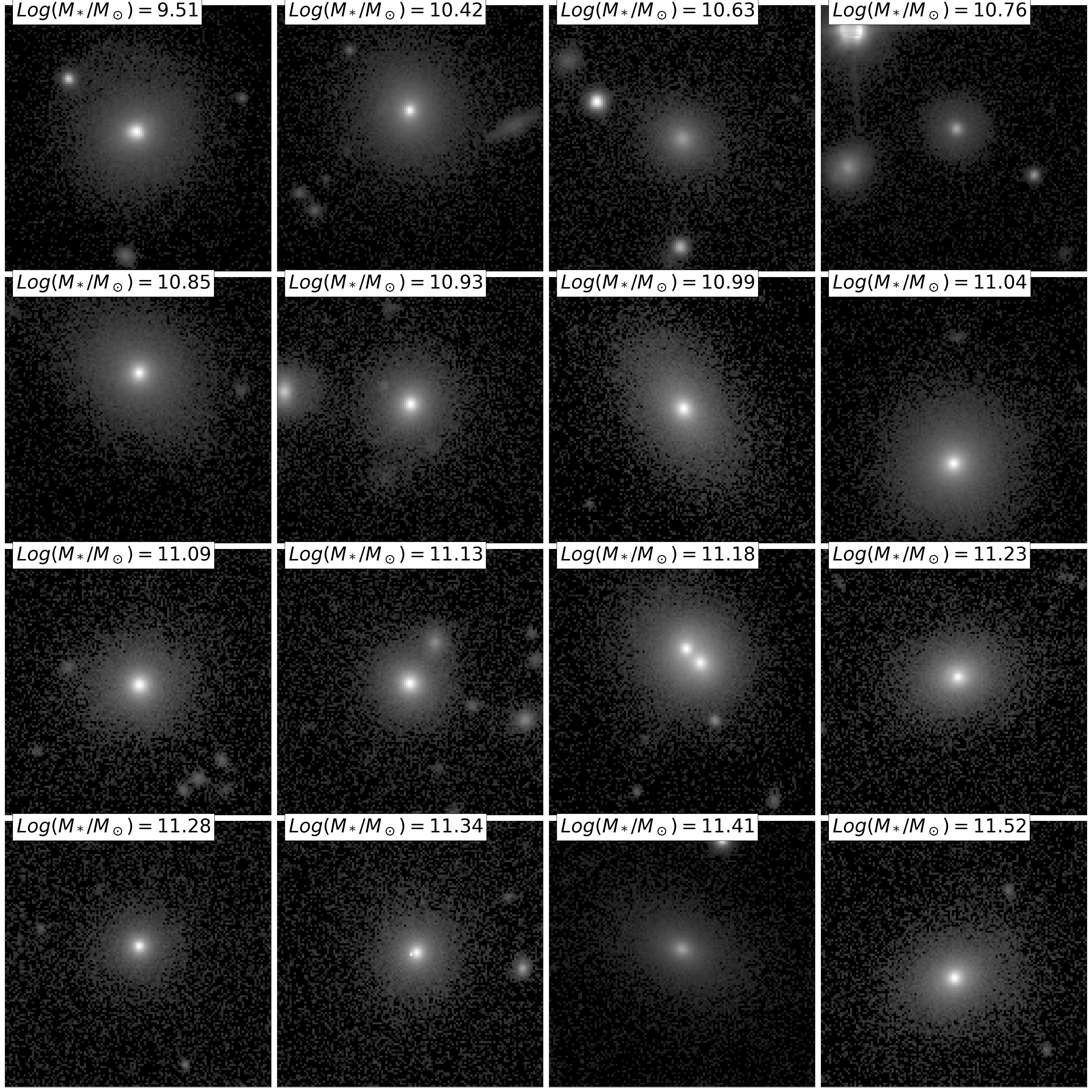}\\
       \hline
       \includegraphics[width=0.45\textwidth]{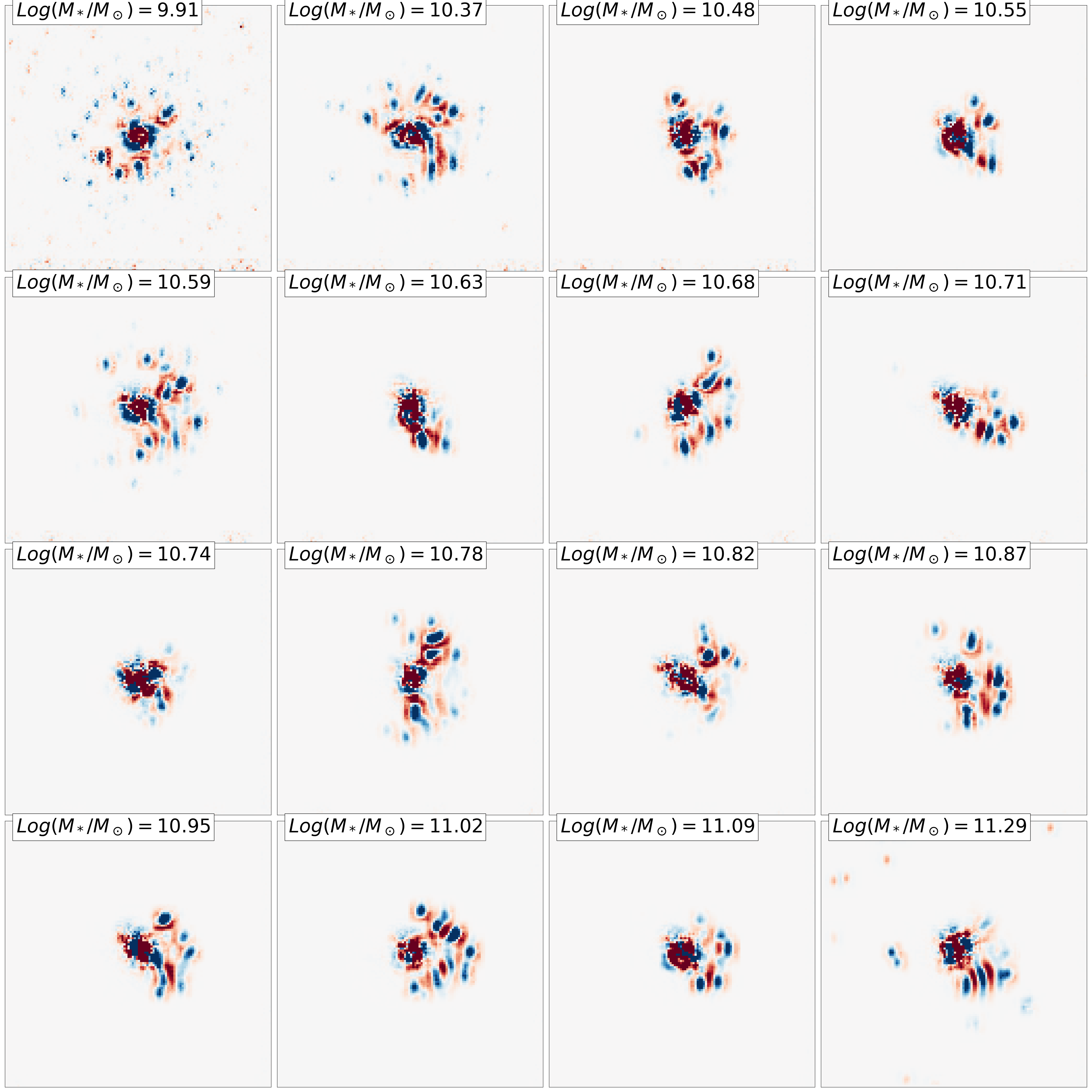} & \includegraphics[width=0.45\textwidth]{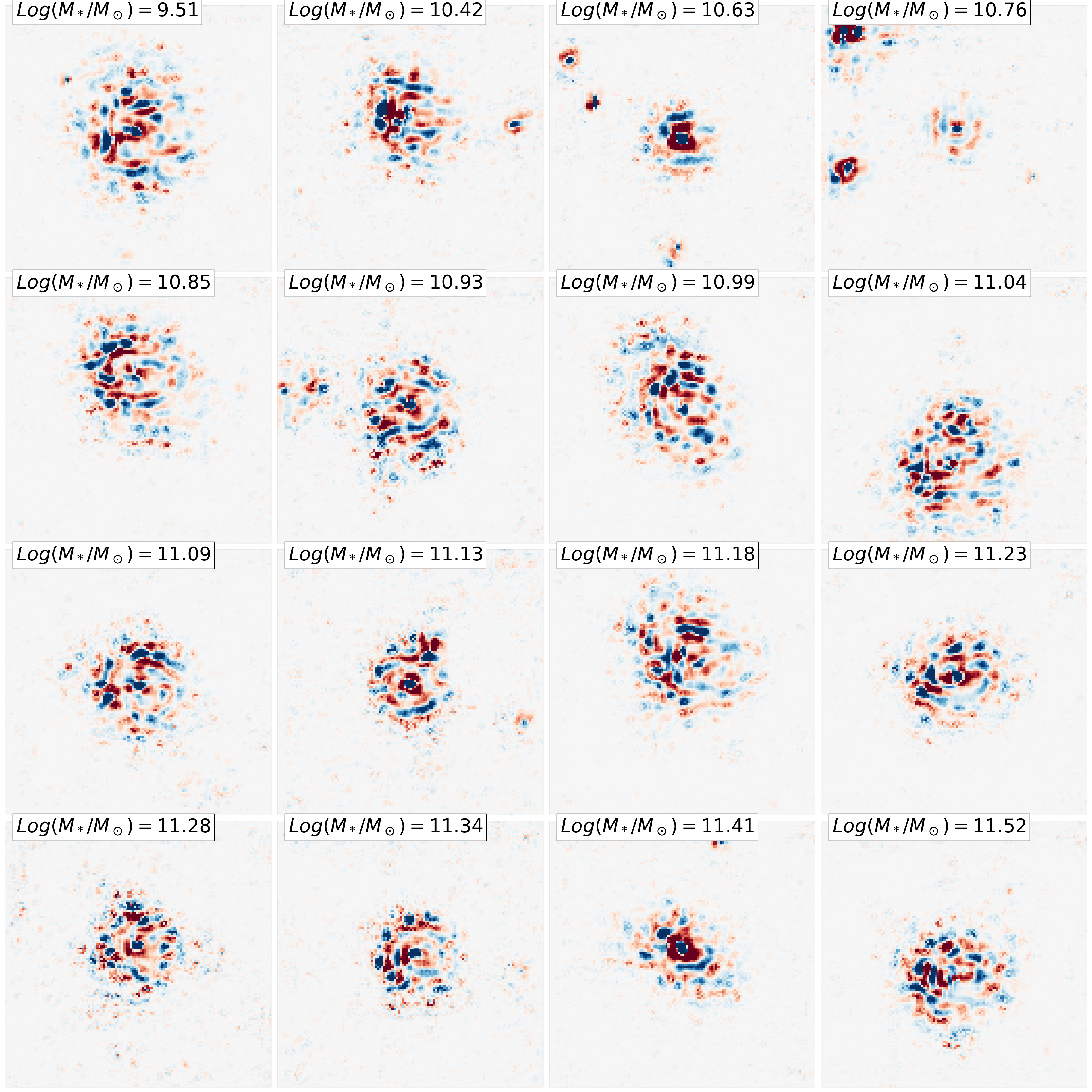}\\
      {\small IllustrisTNG Ellipticals} &  {\small SDSS Ellipticals} \\
     \hline
\end{tabular}
\caption{Example stamps of galaxies classified as Es (ellipticals) by the CNN sorted by increasing stellar mass.  The left panel shows simulated galaxies and the right panel observed ones. The stamps are all $128\times128$ pixels ($50^{"}$ FoV). For visualization purposes, images have been normalized and converted to png with a non linear hyperbolic sine normalization to better appreciate the outskirts. The gray scale is arbitrary. The bottom panels show the attribution maps of the same images computed through integrated gradients. Blue and red colors indicate negative and positive values respectively. No response (0 values) are represented in white. The maps are normalized between the maximum and minimum values so the units are arbitrary and they only reflect variations from a blank image.}
\label{fig:stamps_Es}
\end{figure*}

\begin{figure*}
\begin{tabular}{|c|c|}
      \hline
      \includegraphics[width=0.45\textwidth]{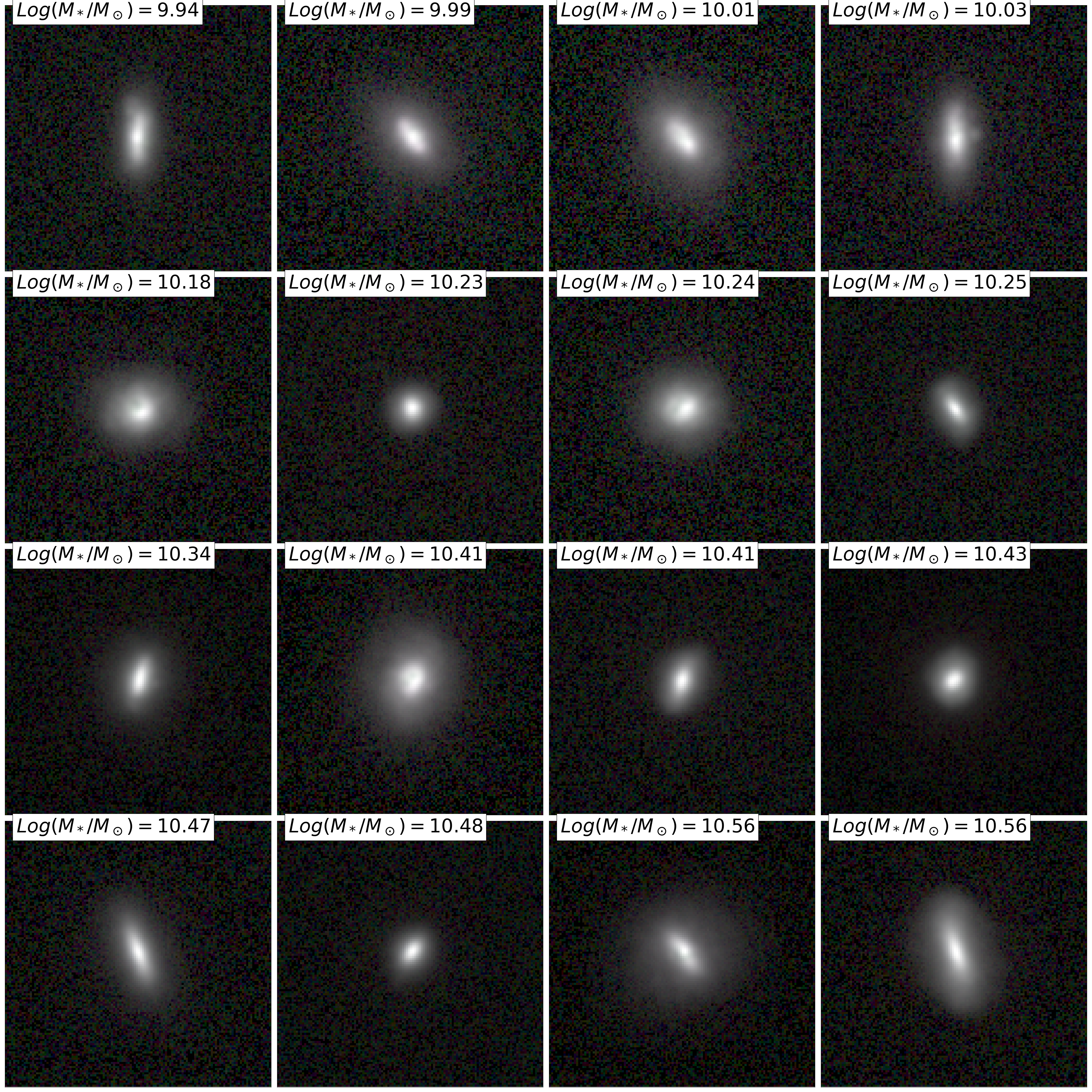} & \includegraphics[width=0.45\textwidth]{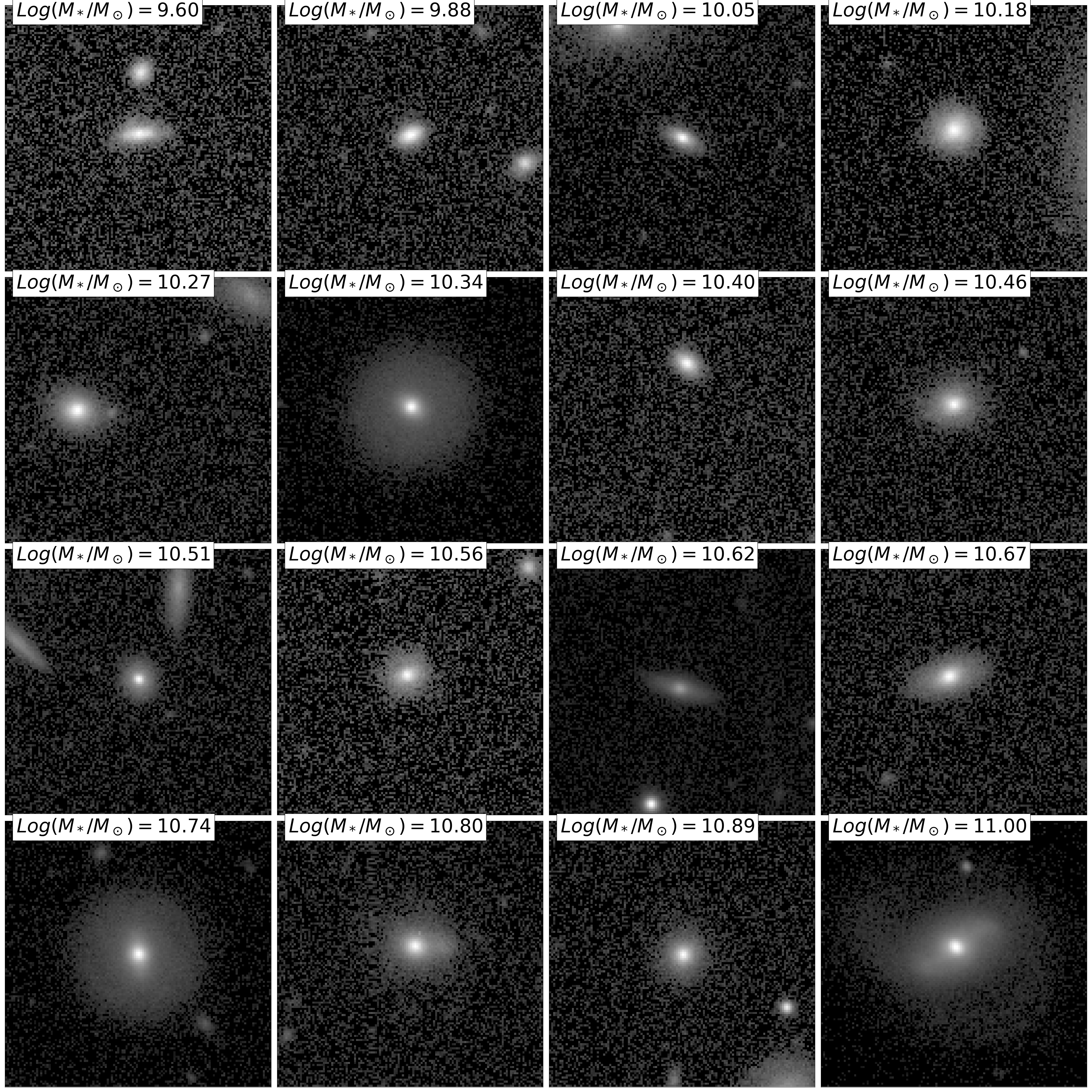}\\
      \hline
       \includegraphics[width=0.45\textwidth]{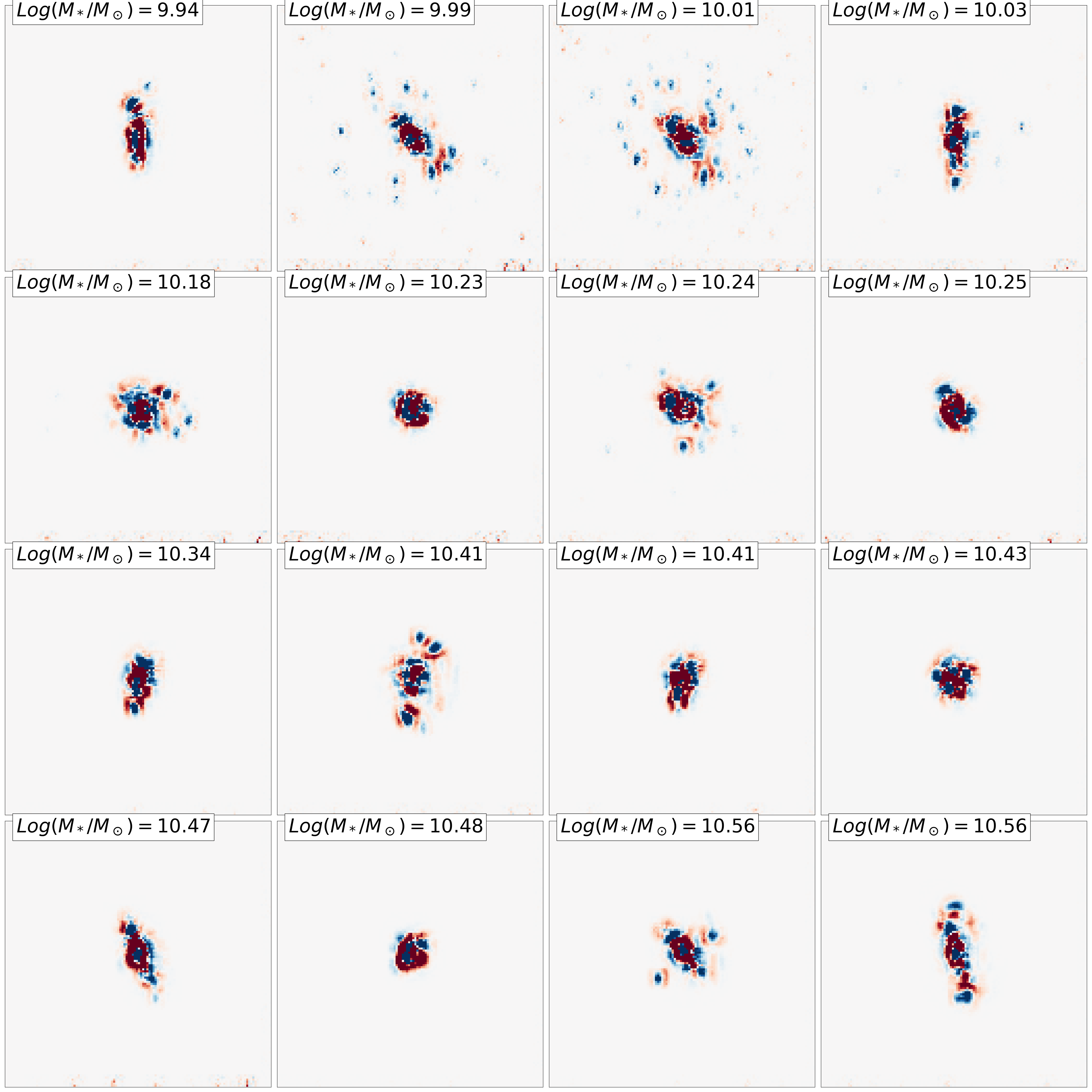} & \includegraphics[width=0.45\textwidth]{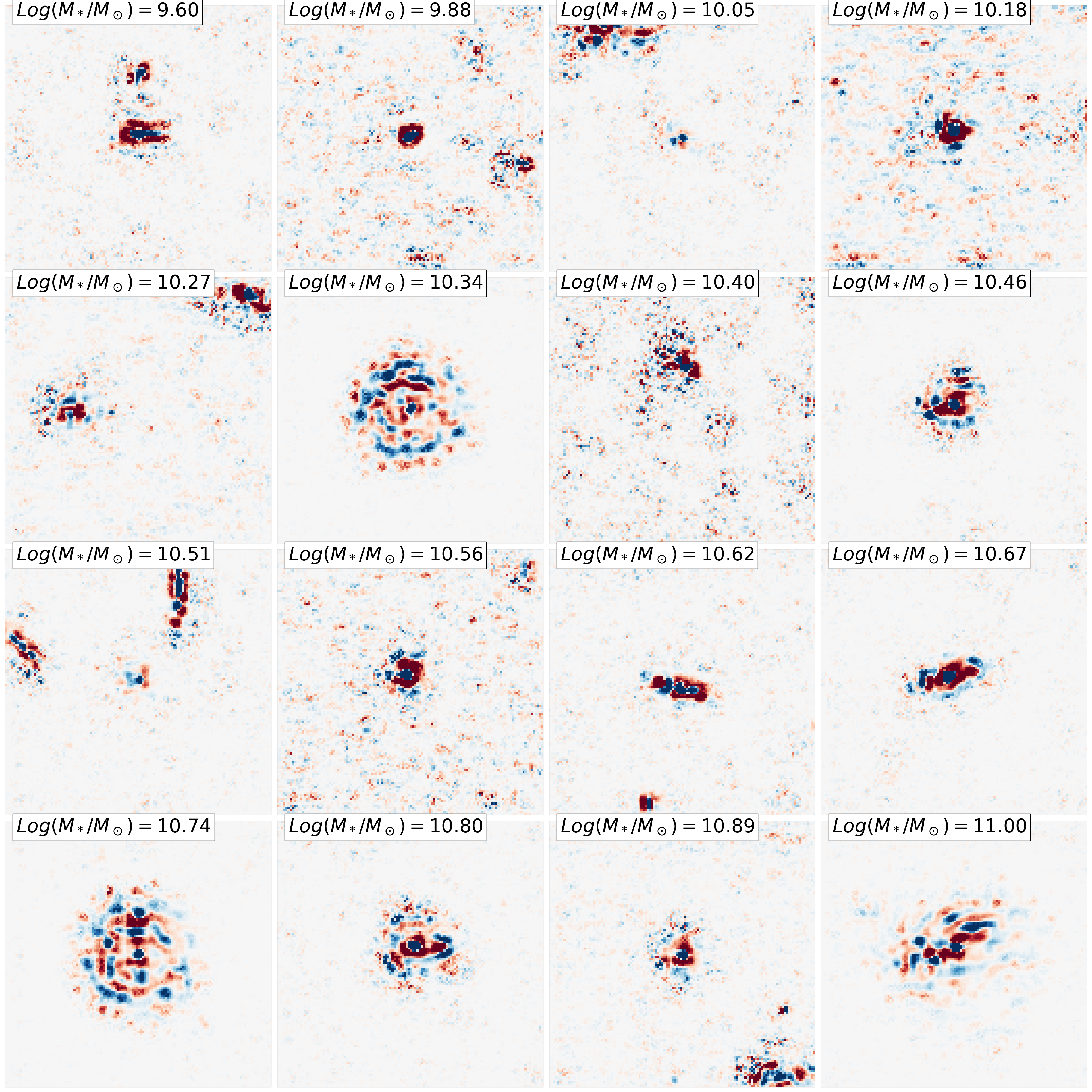}\\
      {\small IllustrisTNG S0s} &  {\small SDSS S0s} \\
      \hline
\end{tabular}
\caption{Example stamps of galaxies classified as S0s (lenticulars) sorted by increasing stellar mass. The left panel shows simulated galaxies and the right panel observed ones. The stamps are all $128\times128$ pixels ($50^{"}$ FoV). Images have been converted to png with a non linear hyperbolic sine normalization to better appreciate the outskirts. The gray scale is arbitrary. The bottom panels show the attribution maps of the same images computed through integrated gradients. Blue and red colors indicate negative and positive values respectively. No response (0 values) are represented in white. The maps are normalized between the maximum and minimum values so the units are arbitrary and they only reflect variations from a blank image.}
\label{fig:stamps_S0s}
\end{figure*}

\begin{figure*}
\begin{tabular}{|c|c|}
      \hline
      \includegraphics[width=0.45\textwidth]{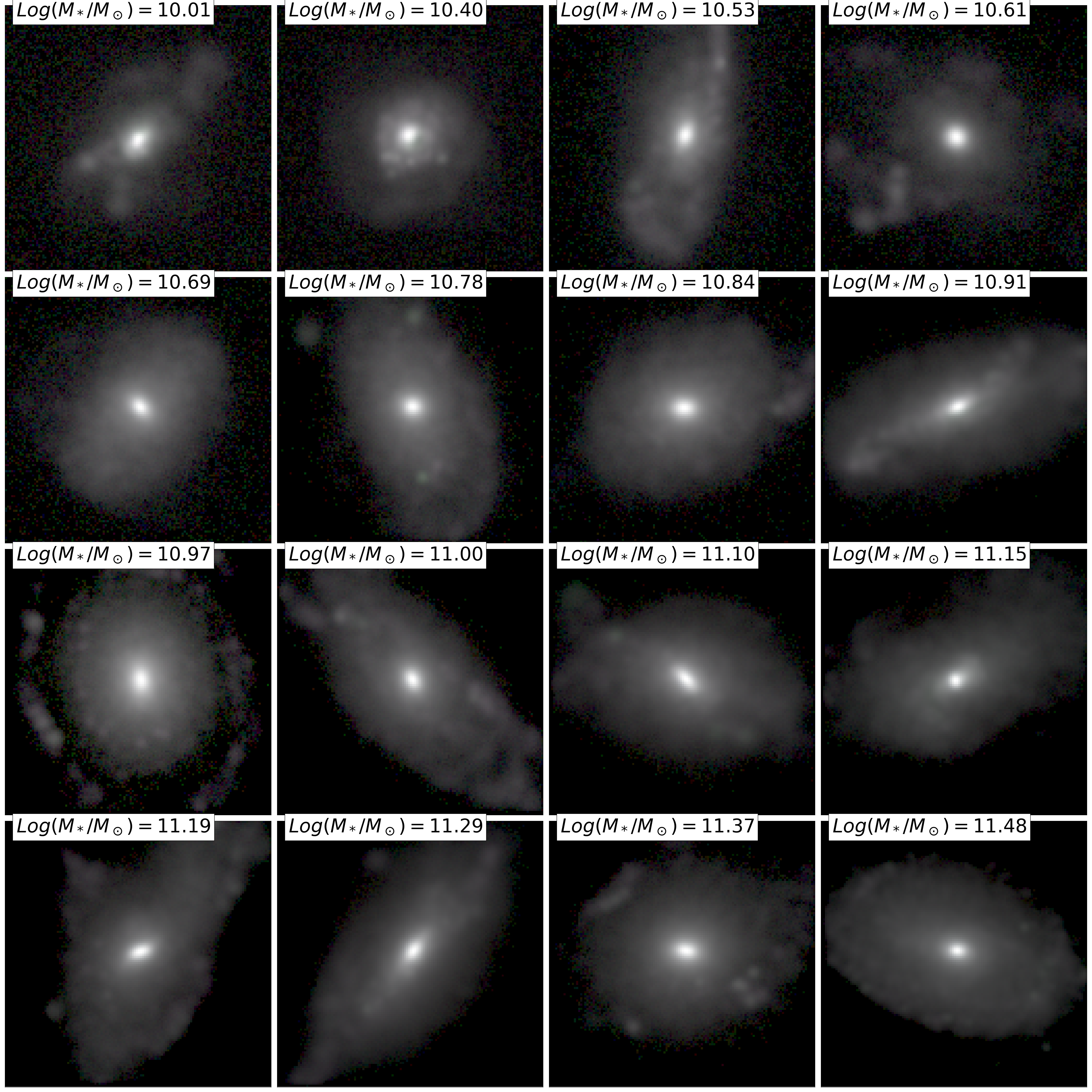} & \includegraphics[width=0.45\textwidth]{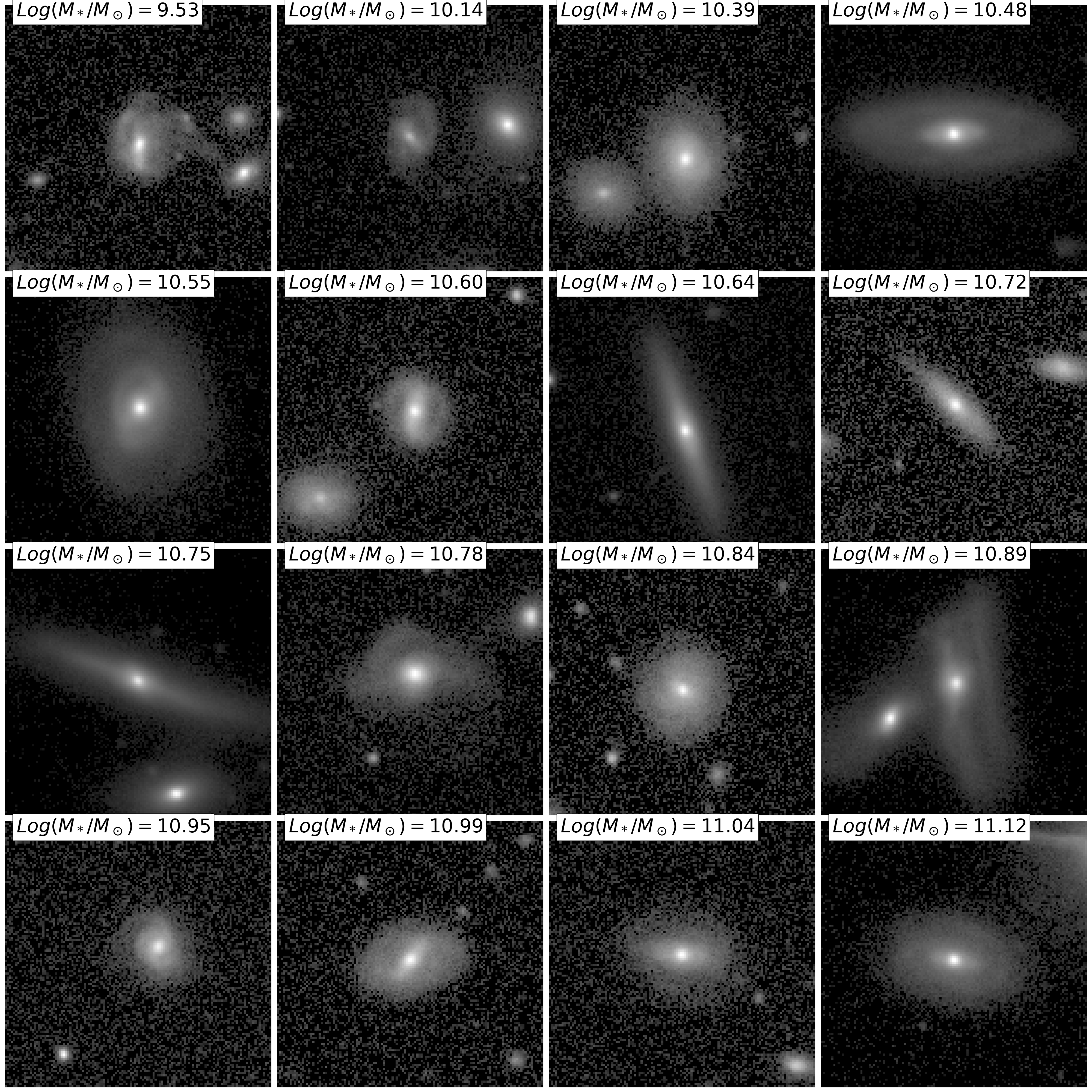}\\
      \hline
      \includegraphics[width=0.45\textwidth]{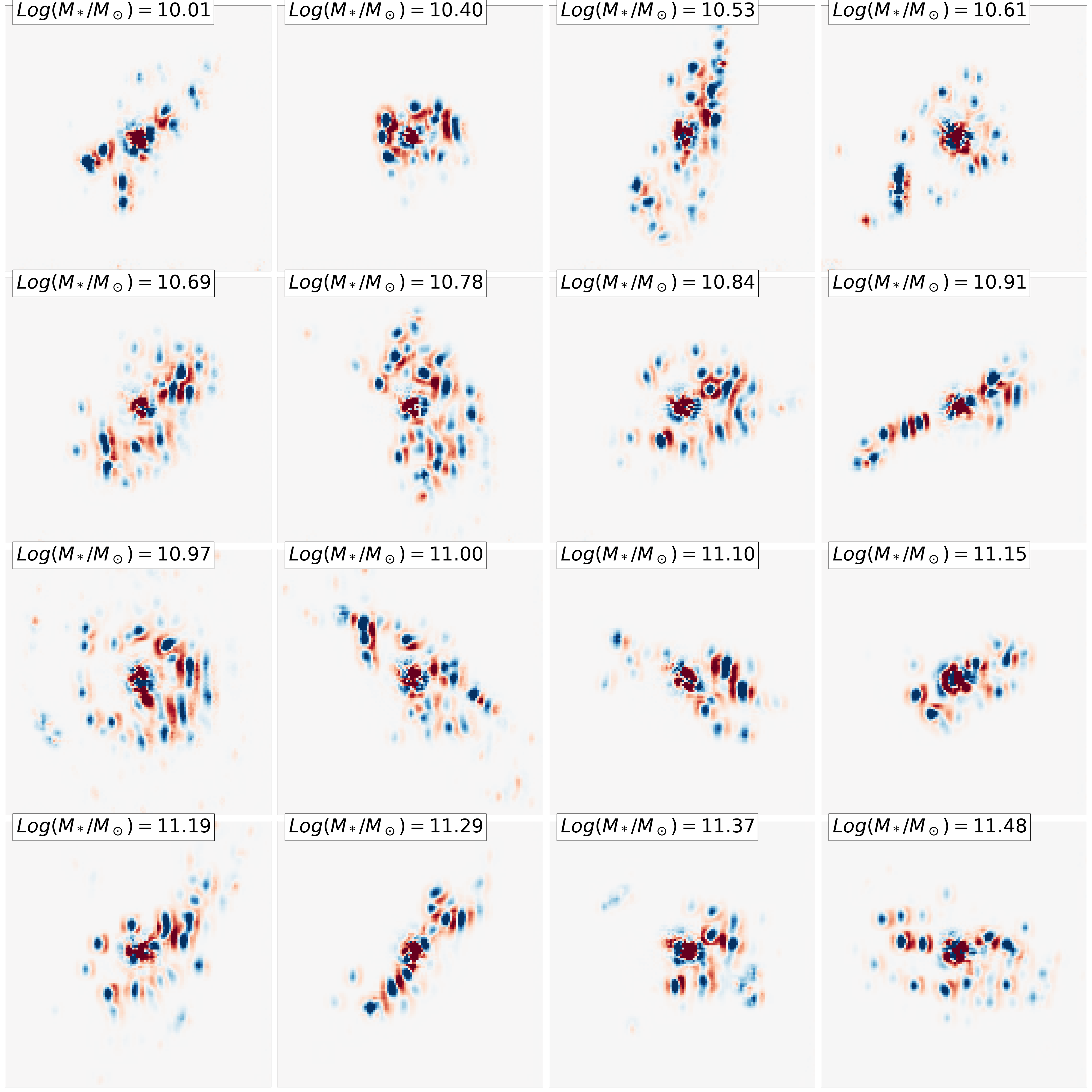} & \includegraphics[width=0.45\textwidth]{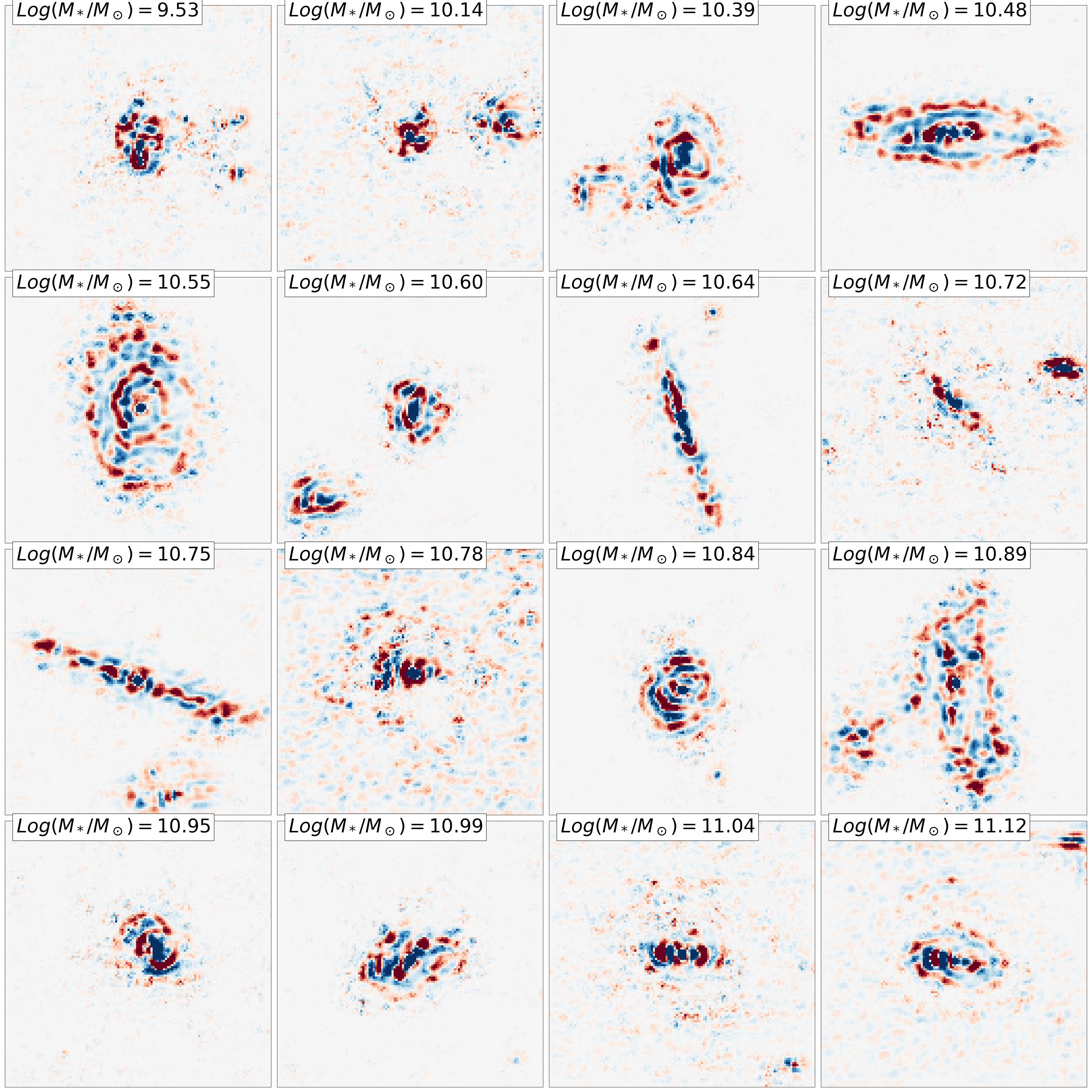}\\
     
      {\small IllustrisTNG Sabs} &  {\small SDSS Sabs} \\
      \hline
\end{tabular}
\caption{Example stamps of galaxies classified as Sabs (early-type spirals) sorted by increasing stellar mass. The left panel shows simulated galaxies and the right panel observed ones. The stamps are all $128\times128$ pixels ($50^{"}$ FoV). Images have been converted to png with a non linear hyperbolic sine normalization to better appreciate the outskirts. The gray scale is arbitrary. The bottom panels show the attribution maps of the same images computed through integrated gradients. Blue and red colors indicate negative and positive values respectively. No response (0 values) are represented in white. The maps are normalized between the maximum and minimum values so the units are arbitrary and they only reflect variations from a blank image.}
\label{fig:stamps_Sabs}
\end{figure*}

\begin{figure*}
\begin{tabular}{|c|c|}
      \hline
      \includegraphics[width=0.45\textwidth]{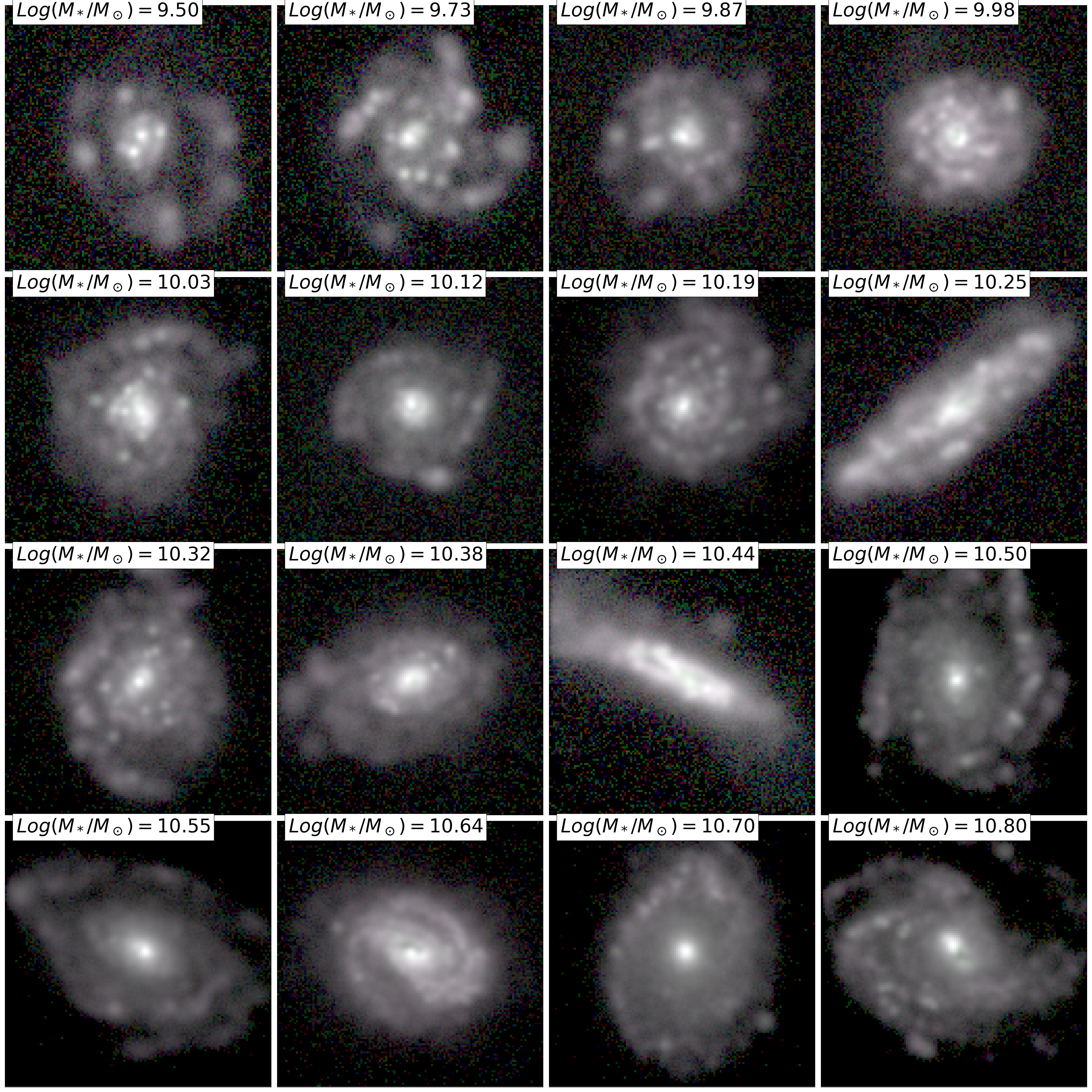} & \includegraphics[width=0.45\textwidth]{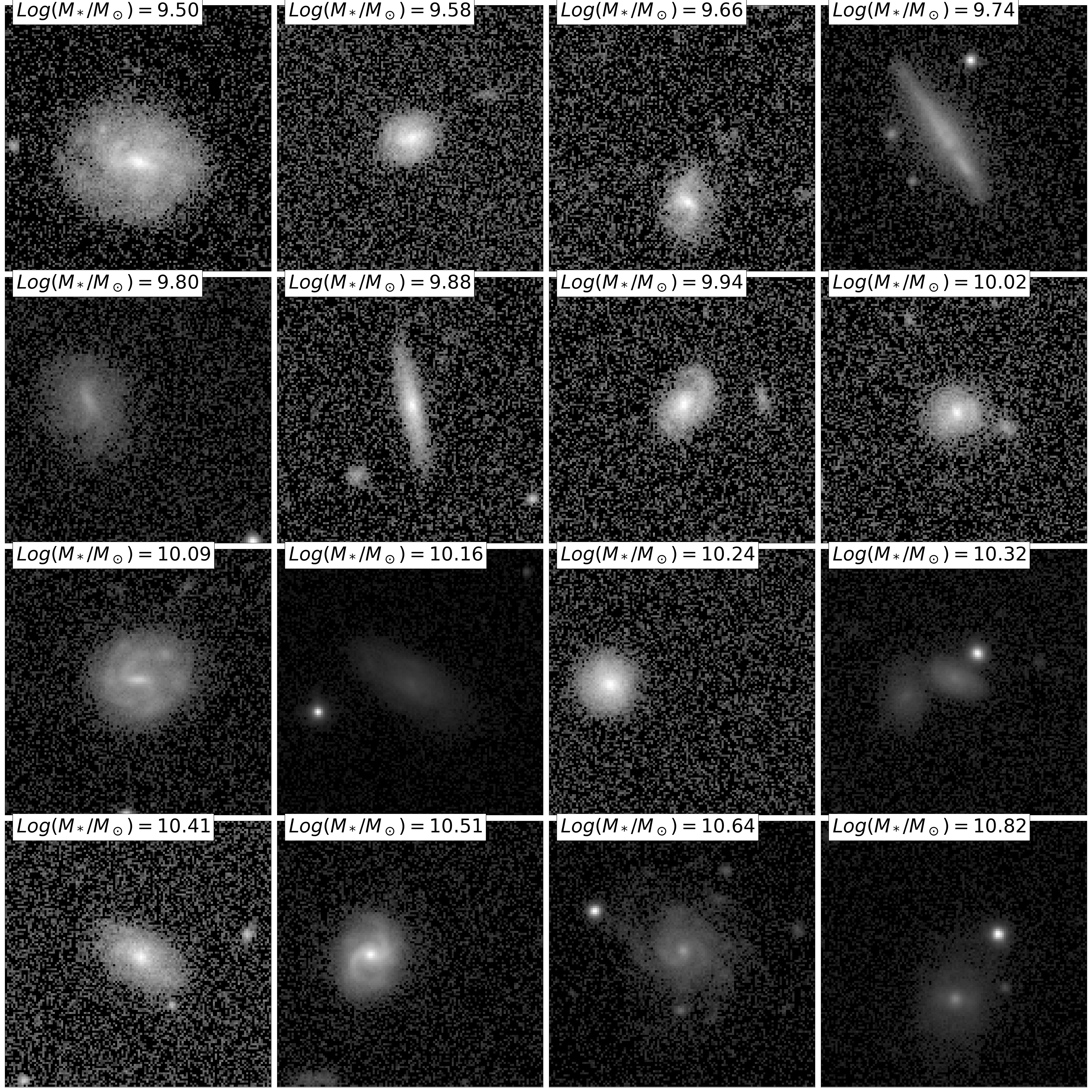}\\
      \hline
      \includegraphics[width=0.45\textwidth]{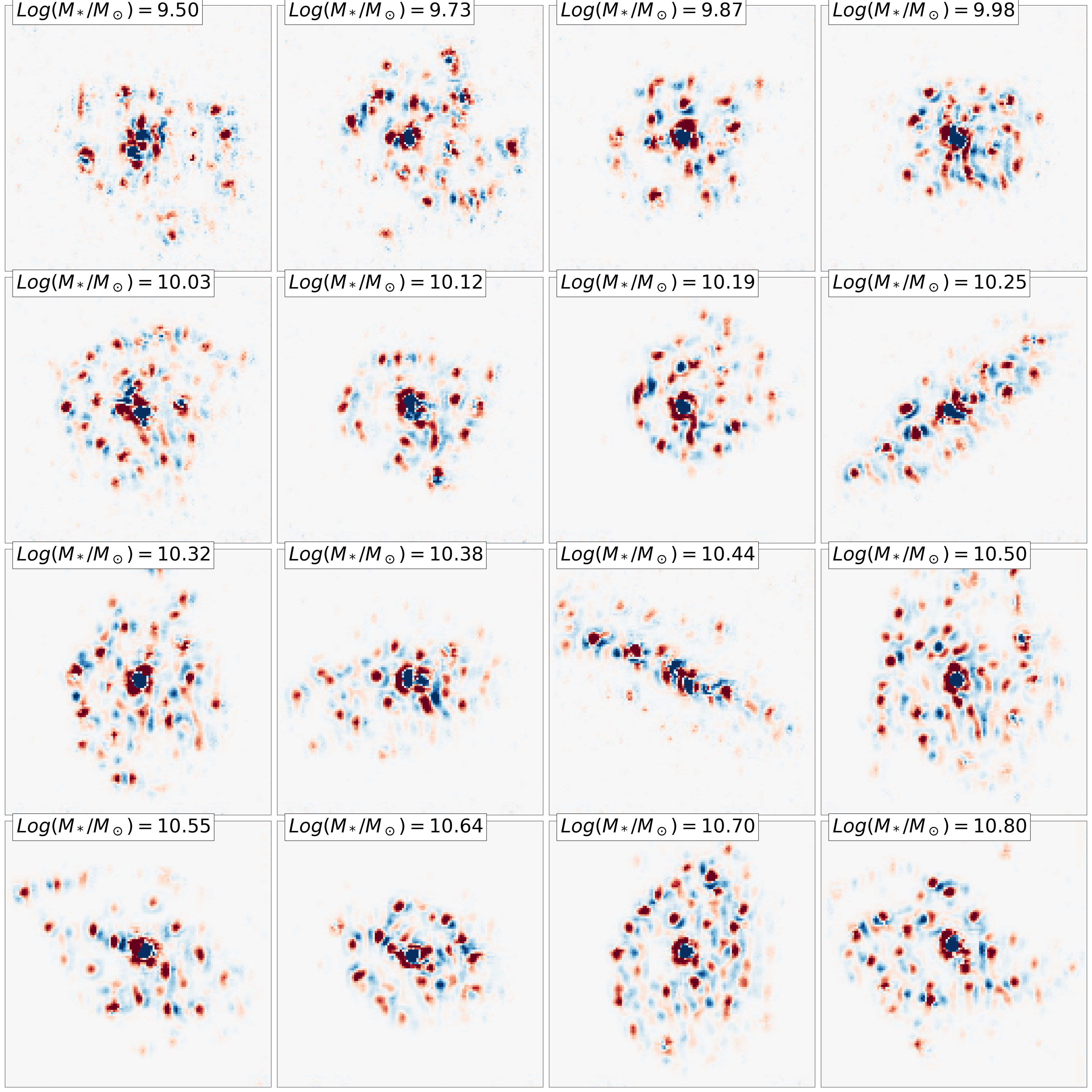} & \includegraphics[width=0.45\textwidth]{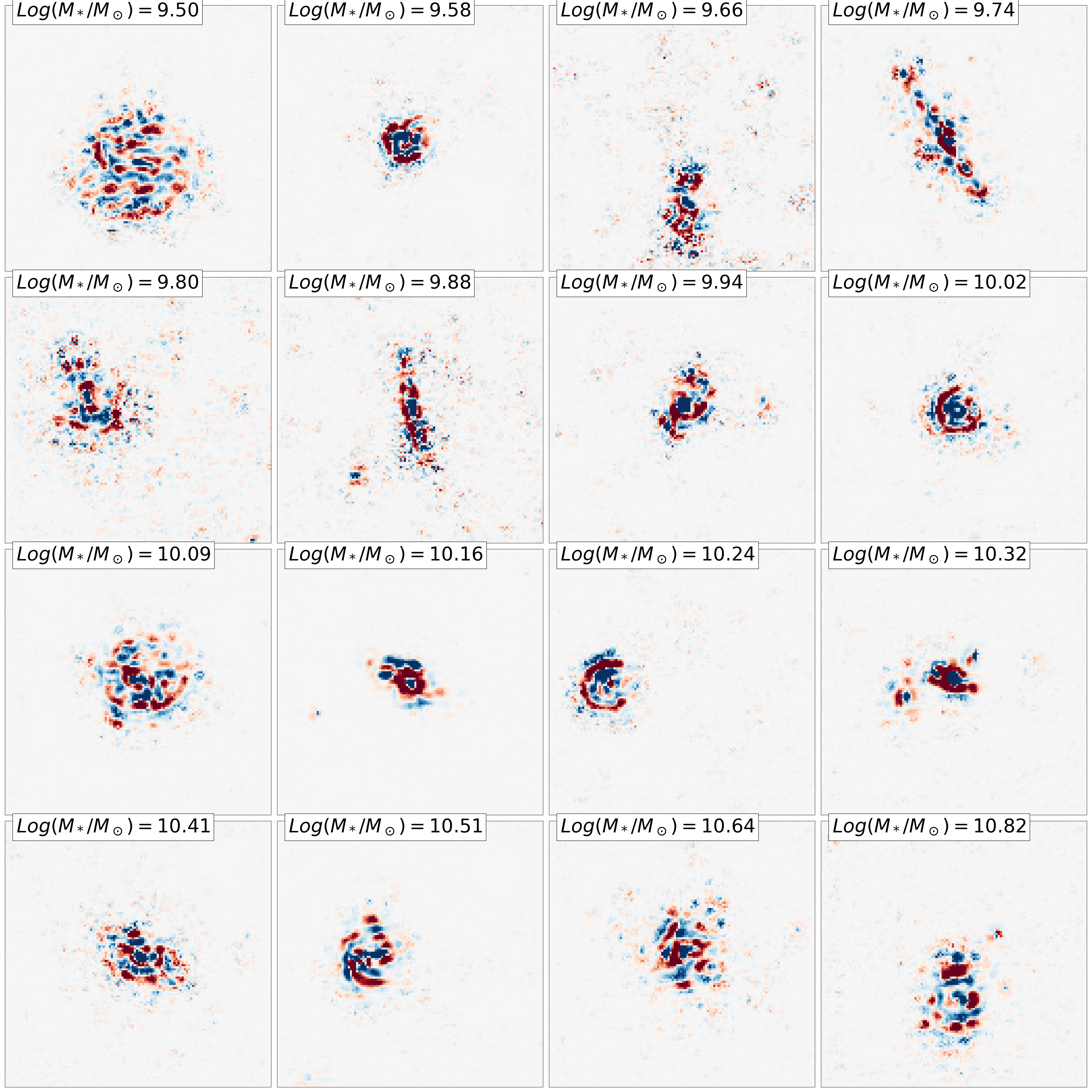}\\
      {\small IllustrisTNG Scds} &  {\small SDSS Scds} \\
      \hline
\end{tabular}
\caption{Example stamps of galaxies classified as Scds (late-type spirals) sorted by increasing stellar mass. The left panel shows simulated galaxies and the right panel observed ones. The stamps are all $128\times128$ pixels ($50^{"}$ FoV). Images have been converted to png with a non linear hyperbolic sine normalization to better appreciate the outskirts. The gray scale is arbitrary. The bottom panel show the attribution maps of the same images computed through integrated gradients. Blue and red colors indicate negative and positive values respectively. No response (0 values) are represented in white. The maps are normalized between the maximum and minimum values so the units are arbitrary and they only reflect variations from a blank image.}
\label{fig:stamps_Scds}
\end{figure*}

We then simply put a galaxy in the class of maximum probability. In the following, Early type galaxies include Ellipticals and S0/a's (also called lenticulars) and Late-type galaxies include Sabs and Scds. We will also refer to Sab galaxies as early-type spirals, and to Scd objects as late-type spirals. 
Figures~\ref{fig:stamps_Es} to~\ref{fig:stamps_Scds} show some example stamps of galaxies classified in the 4 types ordered by increasing stellar mass both in the TNG and in the M15 datasets. As can be appreciated, elliptical galaxies are mostly pure dominated bulge systems. S0 or lenticular galaxies present a dominant bulge component but tend to have a disk with no marked features. Sab galaxies (early-type spirals) have smaller but still noticeable bulges and large disks with spiral arms and/or visible structure in the disk. Finally, Scd galaxies (late-type spirials) have a very small bulge or no bulge at all and a clumpy disk component or with irregular morphology.

 The first thing to notice is that the CNN model trained on SDSS successfully identifies galaxies in the TNG simulation in the four morphological types and that simulated and observed galaxies in a given class share some obvious features. We emphasize that this does not mean that simulated and observed galaxies are not distinguishable. The network is forced to put galaxies in any of the four classes by construction. The fact that there are objects in the four classes, means only that some of the  features learned by the networks to identify the different morphologies in the images are present both in the simulations and in the observations. As a matter of fact, figure~\ref{fig:stamps_Scds} clearly reveals some discrepancies between simulated and observed Scd galaxies. Simulated objects appear systematically more extended and also generally more clumpy than observed Scds. The two edge-on systems also appear to be thicker than observed edge-on systems. However, the CNN still finds that the closest morphological type is a late-type spiral.  

\section{How realistic are the TNG morphologies from the machine learning perspective?}
\label{sec:bnns}

The previous section has shown that the CNNs trained on the visual morphologies from the N10 samples find objects in all four classes also in the TNG simulation. One first interesting question is how \emph{confident} the networks are about the classification in the simulation. Machine learning algorithms will always try to associate objects with the classes they were trained with because there is an implicit assumption that there is a perfect match between the training and test datasets. This is not necessarily the case in this work since we are training in the observational domain and inferring in the simulated one. 

In the following we try to quantify the similarity between the simulated and observed morphologies from the neural network perspective. We adopt two different approaches. First we measure the network confidence by inferring the uncertainties  through a bayesian approach. Secondly, we compare the features learned by the CNN in the simulated and observed samples. We stress that this exercise is not probing whether simulated and observed galaxies can be distinguished. 

\subsection{Bayesian Neural Networks}

\begin{figure}
\includegraphics[width=0.45\textwidth]{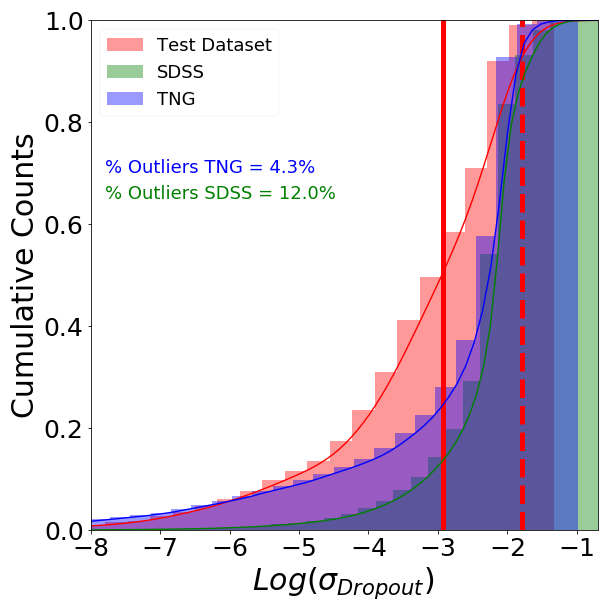}
\caption{Cumulative distribution of the logarithm of uncertainties of the morphological classification estimated through MonteCarlo Dropout. The red histogram shows the distribution for a test set of the N10 sample. The blue histogram corresponds to the TNG sample and the red one is for the M15 dataset. The fraction of outliers defined as the fraction of objects with uncertainties larger than 3 times the standard deviation of the N10 distribution is also indicated.} 
\label{fig:BNN}
\end{figure}

Dropout was first introduced as a method to reduce the risk of overfitting when training deep neural networks \citep{2012arXiv1207.0580H}. By randomly removing some neurons during the training phase, we do not allow neurons to become too specific and ease generalization. \cite{2015arXiv150602142G} have shown that MonteCarlo Dropout in the inference phase can be formally used to approximate the model uncertainty.  We adopt this approach in this work to associate an uncertainty measurement to all classified galaxies. Every galaxy is classified 500 times dropping out a variable fraction of neurons ranging from $30\%$ to $50\%$ at each layer, including the convolutional layers. That way, instead of having a single softmax probability value, every galaxy has an associated probability distribution, arising from the 500 classifications. We then use the standard deviation of the distribution for each galaxy as an estimator of the model uncertainty. We compute the uncertainties in three different samples. First on the fraction of the N10 set which was not used for training (test set). This should be considered as the best case since it has the same properties as the training set. Uncertainties are also computed on the complete M15 dataset (see section~\ref{sec:obs}) and in the TNG one. The cumulative distributions of uncertainties for the three datasets are shown in figure~\ref{fig:BNN}. As expected, the test dataset presents on average lower uncertainties. We use this dataset as a reference to define objects with large uncertainties in the SDSS and TNG datasets. To that purpose we compute the median error value and the standard deviation of the distribution of uncertainties in the test dataset and define as outliers objects with a measured uncertainty larger than 3 times the standard deviation. 

\begin{figure*}
\begin{tabular}{|c|c|}
      \hline
      \includegraphics[width=0.45\textwidth]{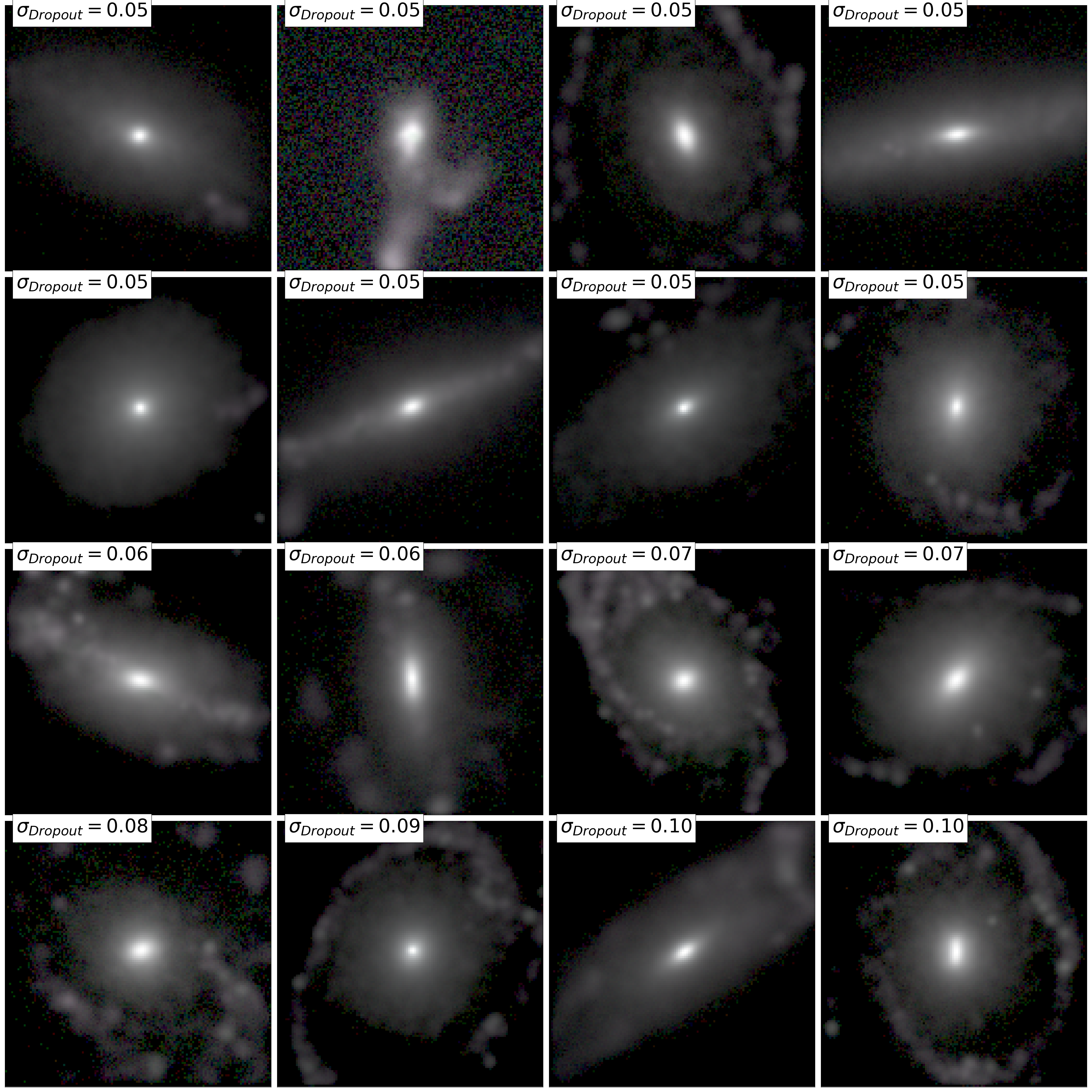} & \includegraphics[width=0.45\textwidth]{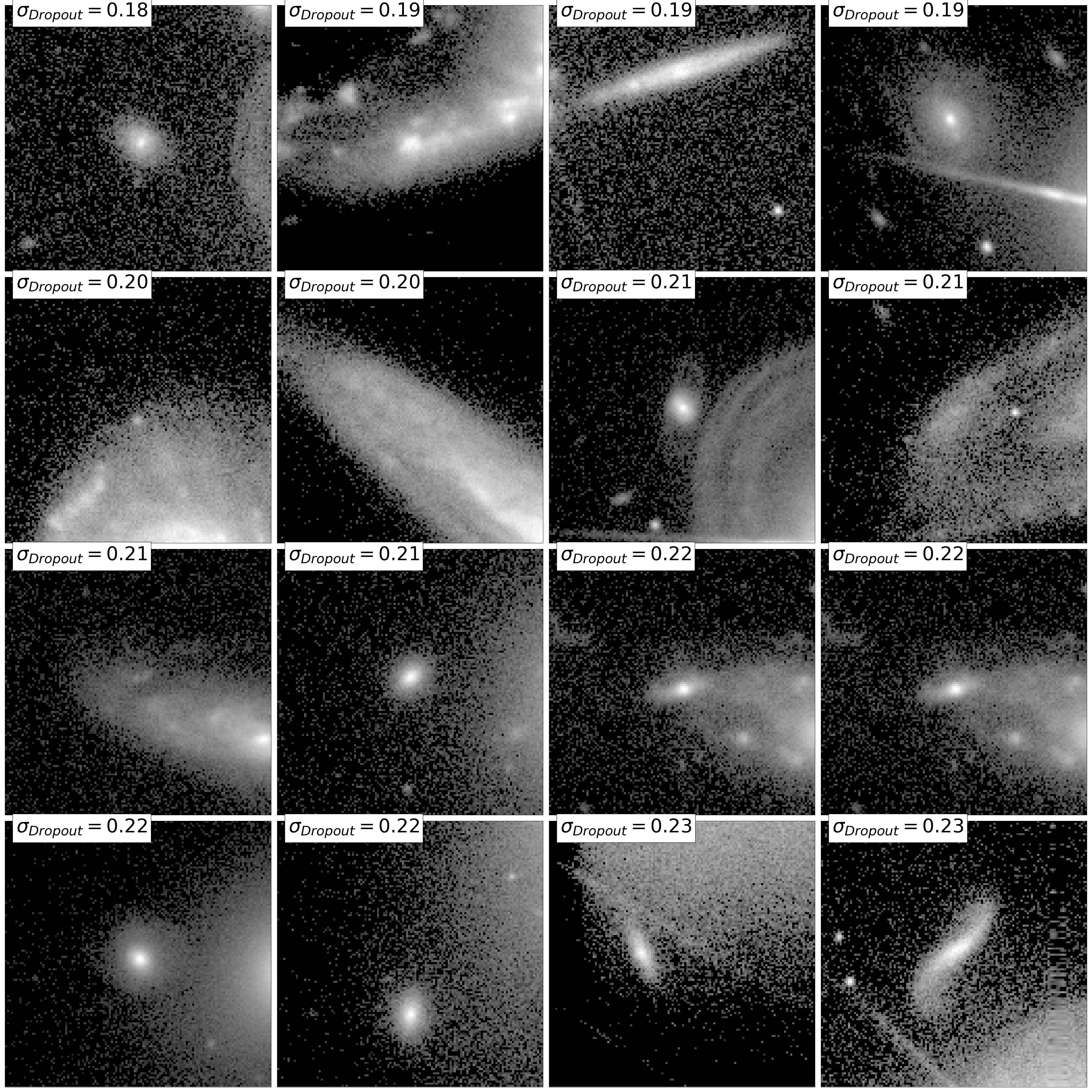}\\
      {\small TNG} &  {\small SDSS} \\
      \hline
\end{tabular}
\caption{Objects with large uncertainty as inferred from the MonteCarlo Dropout technique used in this work. The left (right) panel shows the 16 stamps with larger uncertainty values in the TNG (M15) sample. See text for details. }
\label{fig:outliers}
\end{figure*}

We find that the number of outliers defined that way is $\sim4\%$ for TNG and $\sim12\%$ for the M15 dataset. The first point we can conclude is that the fraction of simulated galaxies with more uncertain classification as compared to the reference observational sample is only$\sim4\%$. This confirms the realism of galaxy morphologies in the TNG run or at least that the features learned by the CNNs to identify the different morphologies are found in both the simulations and the observations. One interesting question that arises is what are these features. We try to address this in the following section by exploring the attribution maps. In the left panel of figure~\ref{fig:outliers} we show the images of the 16 galaxies in TNG with largest uncertainties. A distinguishing feature for many of them is a ring of clumpy star formation around the galaxies. This feature is apparently not seen in the observations at least around early-type galaxies which can explain the high uncertainty values. It confirms that the MonteCarlo Dropout technique is able to identify objects with peculiar features. 

A surprising result is that the fraction of outliers is larger in the M15 sample than in the simulations. Since the CNN was trained with a sample from SDSS, one would have naturally expected that the M15 dataset and the N10 sample used for training should be closer. In the right panel of figure~\ref{fig:outliers} we explore the objects in the M15 sample with larger error values. In most of the cases they are small galaxies with very bright companions. These objects are less abundant in the training set since a clean sample of very nearby galaxies was selected by \cite{2010ApJS..186..427N}. Also the simulations do not have companions by construction. Another element which can help explaining the larger outlier fraction in the M15 dataset is the fact that it extends to fainter magnitudes than the N10 sample used for training. The lower SNR can thus increase the model uncertainty. This explains the larger uncertainty values in the SDSS. It also confirms that our error measurement is sensitive to outliers. 

\subsection{Comparison of attribution maps and features learned}

Deep neural networks have been proven to produce very high classification accuracies. The price to pay is less control on the features used by the networks to perform the classification. In this section we try to explore the similarity between the features extracted by the network in the simulations and in the observations as a way to quantify how close are the morphologies between TNG and SDSS. Figures~\ref{fig:stamps_Es} to~\ref{fig:stamps_Scds} show the attribution maps for the same galaxies represented in the top panels. The maps indicate the pixels that contributed most to the network decision for a given image. They are computed here using Integrated Gradients~\citep{2017arXiv170301365S}, but other attribution techniques we tested gave similar results.  The maps are directly not translatable into a \emph{physical} set of features but can help localizing where in the galaxies is the information used for classification. We do see that the attribution maps generally trace the pixels belonging to the galaxy, confirming that the network is ignoring the noise when classifying galaxies. This might appear like a trivial statement, but it is not given that there might be some S/N trends in the training set. For example, Scd galaxies tend to be fainter than Ellipticals so the networks might have learned that a lower S/N is correlated with the morphological type. The attribution maps show it is not the case. We do also observe that for elliptical galaxies the important pixels seem to be more concentrated towards the central regions than for later types which indicates that the network is focusing on the bulge component for these systems as one would expect. 

The comparison between the maps in TNG and SDSS qualitatively reveals that similar features are found in both datasets at fixed morphology.  For Sab galaxies for example (figure~\ref{fig:stamps_Sabs}) we appreciate how the activation pixels similarly trace the disk region. For elliptical galaxies it appears that the attribution maps in the observations look more scattered than in the simulations. This is likely due to the fact that ellipticals in SDSS tend to be in over dense regions with satellites which is not the case in the TNG images since only the central galaxy was rendered. In summary the maps provide limited information (or at least not easily interpretable in terms of physical quantities measured in images) but allow one to check that no major differences are found between the simulated and observed samples.

\begin{figure}
\includegraphics[width=0.45\textwidth]{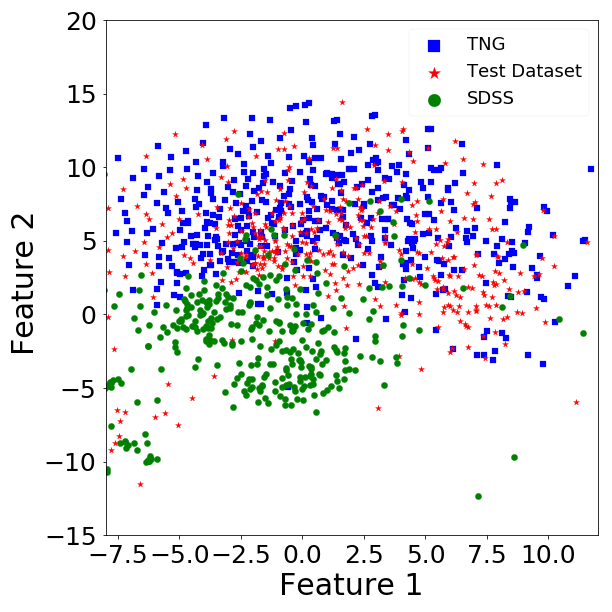}
\caption{Visualization of the features learned by the CNN in a 2-dimensional space obtained with tSNE (see text for deails). The red points show the N10 sample, the green points are galaxies in the M15 dataset and the blue points are simulated galaxies from TNG. }
\label{fig:tSNE}
\end{figure}

In order to better understand if the CNN appreciates noticeable differences between the TNG and SDSS datasets we perform an exploration of the features learned by the network. We extract the features after the last convolutional layer before the dense part of the network and compare them. To that purpose we create a feature vector for a subsample of 500 galaxies from the N10 sample used for training, 500 additional galaxies from the M15 dataset and 500 galaxies from TNG. Since the feature space is highly dimensional ($\sim100,000$) we project it into a 2-dimensional space for visualization purposes using the dimensionality reduction algorithm tSNE (t-Distributed Stochastic Neighbor Embedding, \citealp{2008TsNE}). We use a learning rate of $900$ and a perplexity value of $30$. However changes in these parameters do not change the main trends. The result of this exercise is shown in figure~\ref{fig:tSNE}. Recall that the axes have arbitrary units and do not encapsulate any physical meaning. 
The TNG (blue squares)  and the N10 (red stars) samples form a unique cluster in the space defined by the 2 features extracted with tSNE. It confirms that similar features are found by the CNN in both datasets, which trigger comparable responses of the neurons. Data points from the M15 (green circles) sample are also in the same cluster but seem to be slightly off centered with respect to the 2 other distributions. The reason seems to be that the M10 sample contains fainter galaxies. This is in some sort a limitation of the approach followed in this work since we are using a sample of bright galaxies to train while we are inferring on significantly fainter objects. We will show in the following sections that this does not seem to significantly alter our main conclusions. It confirms however that the visualization of the feature space is sensitive to differences in the galaxy properties. The main conclusion from this plot is therefore that simulated galaxies do not show significant differences in the feature space. Another interesting feature in figure~\ref{fig:tSNE} is that there is a small offset cluster of points containing both M15 and N10 galaxies. A visual inspection of some of the images in the cluster reveals that it is made of galaxies with a bright point source in the stamp. It is not straightforward to explain why these images create a separated cluster. One possibility is that since CNNs do not care about the position of an object in the image (they are by definition translational invariant), a point source could have been interpreted as a bright bulge. In order to avoid this, the network needs to associate some characteristic features for these images.

From these different tests, we can conclude that, at least from the neural network perspective, comparable features are found in the simulations and in the observations. Training on N10 and inferring in TNG seems justified. We can be confident that the morphologies estimated in the TNG sample are reliable. In the next sections, we analyze their physical properties and abundances as compared to observations.

\section{Mass-size relation}
\label{sec:scaling}

We now explore the similarity between simulated and observed galaxies from a more physical perspective, i.e. by looking at the scaling relations. We primarily focus on the stellar mass-size relation. We use the semi major axis of the best Sersic model as a size estimator for galaxies both in the simulations and in the observations. As detailed in section~\ref{sec:obs}, the fitting approach in the M15 sample is fully described in \cite{2015MNRAS.446.3943M}. For the simulations, \cite{2019MNRAS.483.4140R} performed also Sersic fits on the projected 2D light maps that we use here. Although both studies did not use strictly speaking the same method we assume that the one component Sersic fits are stable enough so that no major systematics are introduced. For the stellar mass in the SDSS, we use, for consistency, the stellar mass estimated using the luminosity from the best single Sersic model. In the simulations we use an aperture of $30$ kpc. This aperture has been shown to provide stellar mass estimates in good agreement to those within Petrosian radii in observations~\citep{2015MNRAS.446..521S}and a reasonable compromise for comparison with observations also towards the highest-mass end \citep{2018MNRAS.473.4077P}. As will be shown in section~\ref{sec:SMFs}, the $30$ kpc aperture provides an excellent match to the observed stellar mass function. 

\begin{figure}
\includegraphics[width=0.45\textwidth]{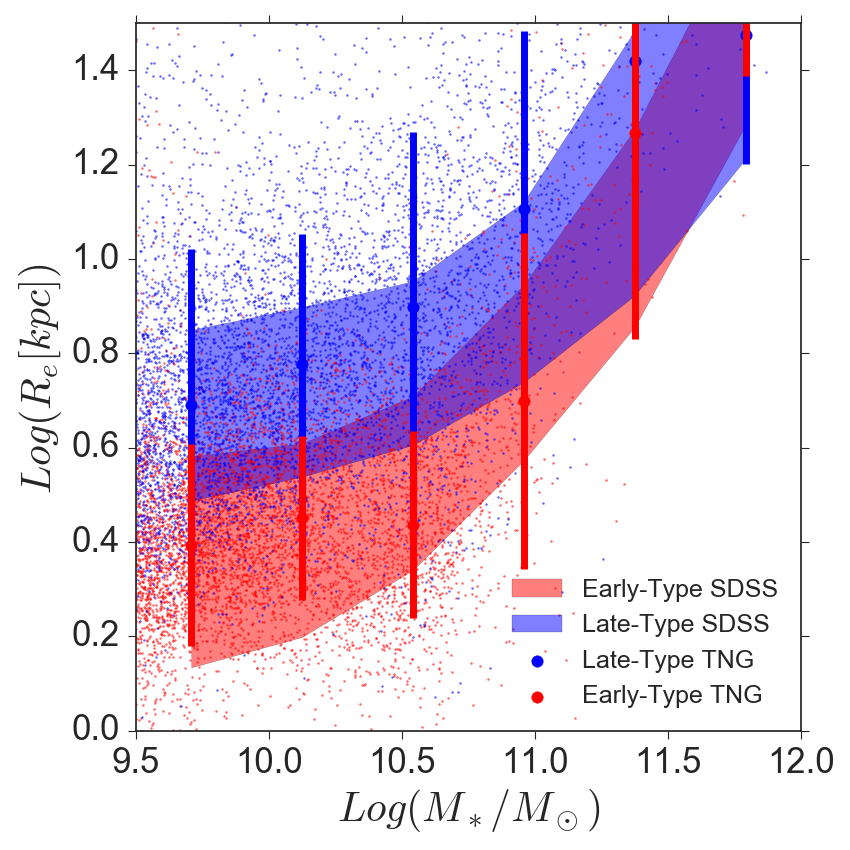}
\caption{Stellar mass - size relation of early-type and late-type galaxies in IllustrisTNG and in SDSS. The shaded red (blue) regions show the observed median mass-size relations along with the scatter for early (late) type galaxies. The red (blue) points show the distribution of individual simulated early (late) type galaxies. The large dots with error bars indicate the medians and scatters in bins of stellar mass respectively.}
\label{fig:mass_size_EL}
\end{figure}

Figure~\ref{fig:mass_size_EL} shows the mass-size relations for early and late-type galaxies. We observe a reasonably good match between observed and simulated galaxies. Namely the simulations reproduce well the largely reported trend in the observations that late-type galaxies are larger than early-type galaxies at fixed stellar mass (e.g. \citealp{2014MNRAS.443..874B}). The slopes of both relations are also well captured in the simulations. Notice that using the same synthetic images, \cite{2019MNRAS.483.4140R} found no significant differences in the sizes of early and late-type galaxies in TNG. This might be due to the fact that they used only the Sersic index to define the two morphological classes while here we are using a definition based on the global appearance of the galaxies. It emphasizes the importance of using accurate global descriptors of morphology. This is a remarkable improvement as compared to the first Illustris run which showed significant discrepancies in the scaling relations (i.e. \citealp{2017MNRAS.467.1033B}), namely shallower slopes and higher normalizations than in the observations. Figure~\ref{fig:mass_size_EL} shows instead that both the slope and the normalization match reasonably well the observations. The simulations present still a slightly larger scatter in size at fixed stellar mass. Note that~\cite{2018MNRAS.474.3976G} also measured a size difference between quenched and star forming galaxies in the TNG simulation (with no morphological selection), also in reasonable agreement with observations not only at z=0 but also to higher redshifts. 

\begin{figure*}
\begin{tabular}{|c|c|}
      \includegraphics[width=0.45\textwidth]{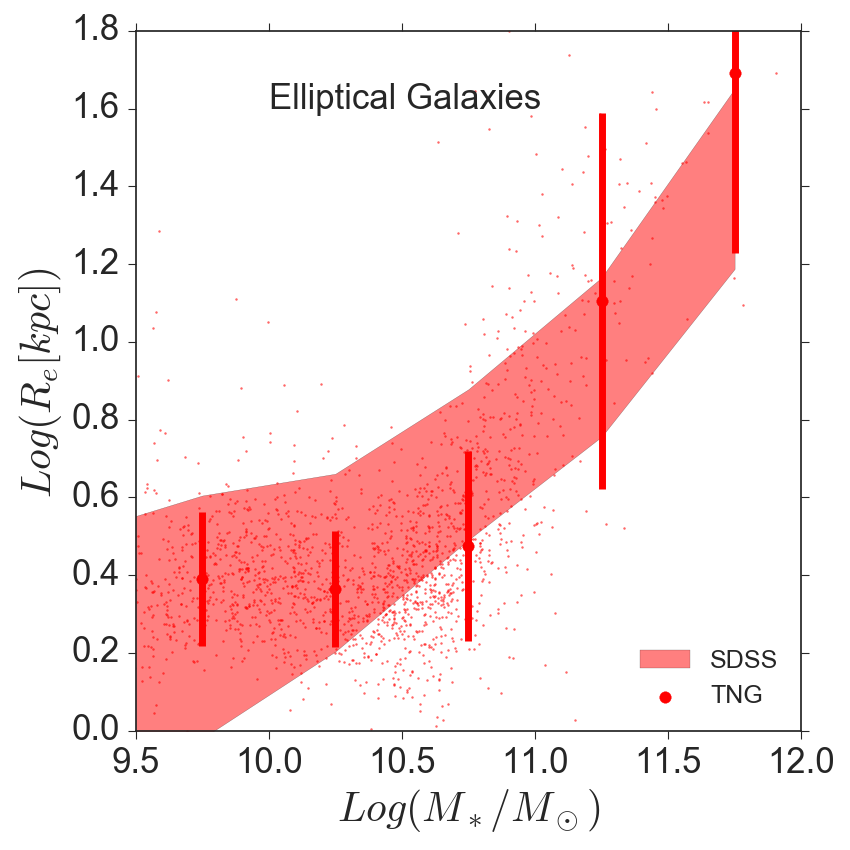} & \includegraphics[width=0.45\textwidth]{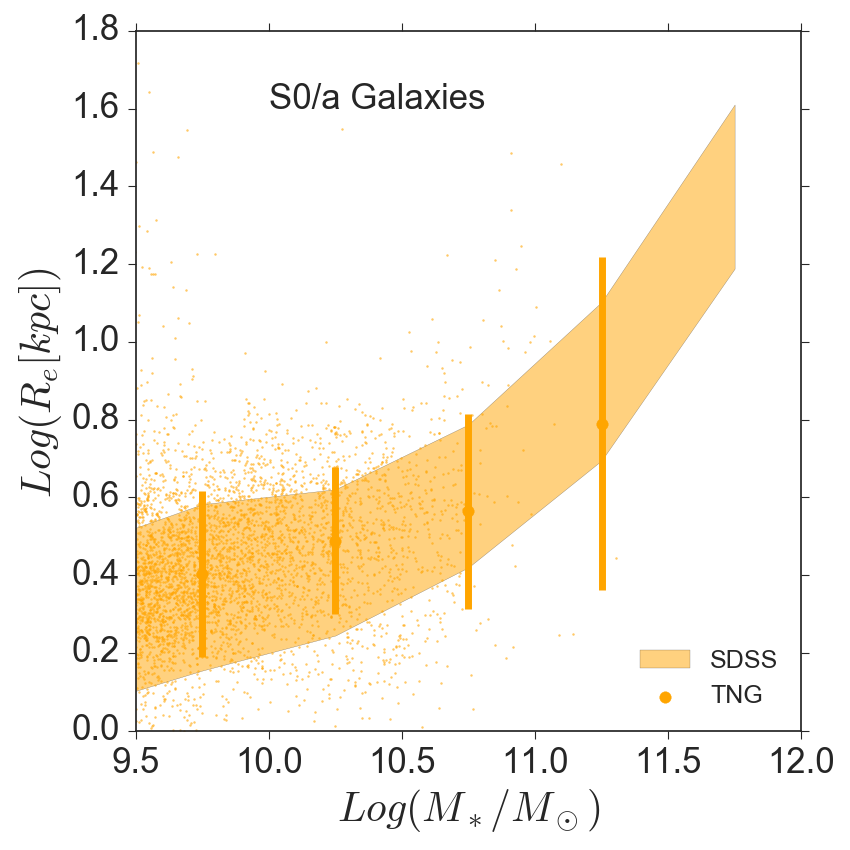}\\
   \includegraphics[width=0.45\textwidth]{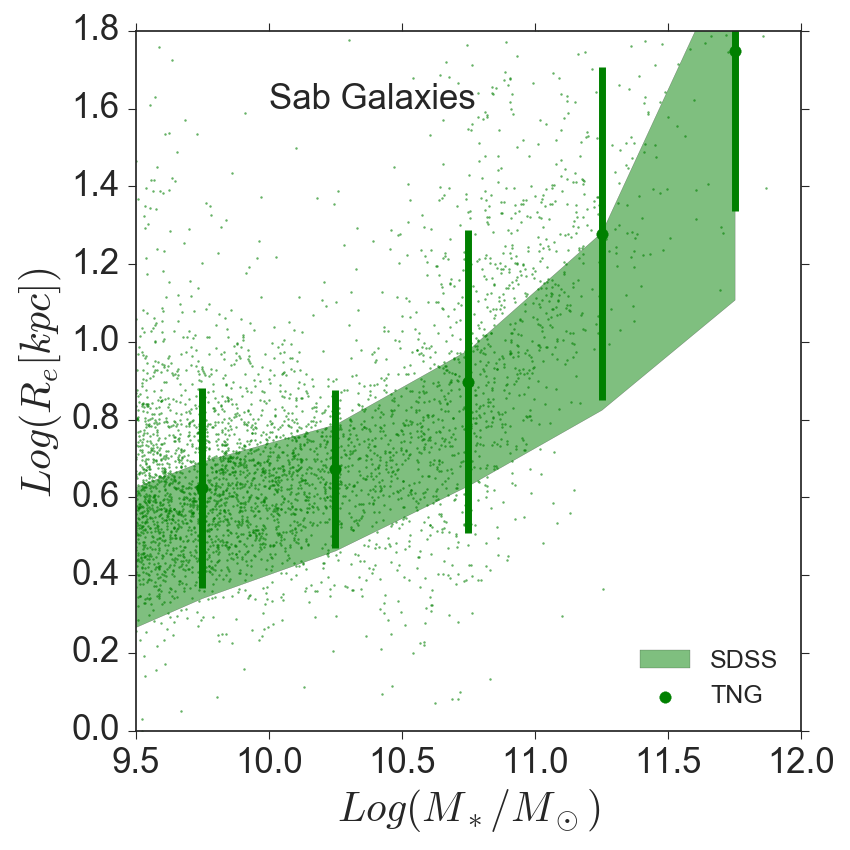} & \includegraphics[width=0.45\textwidth]{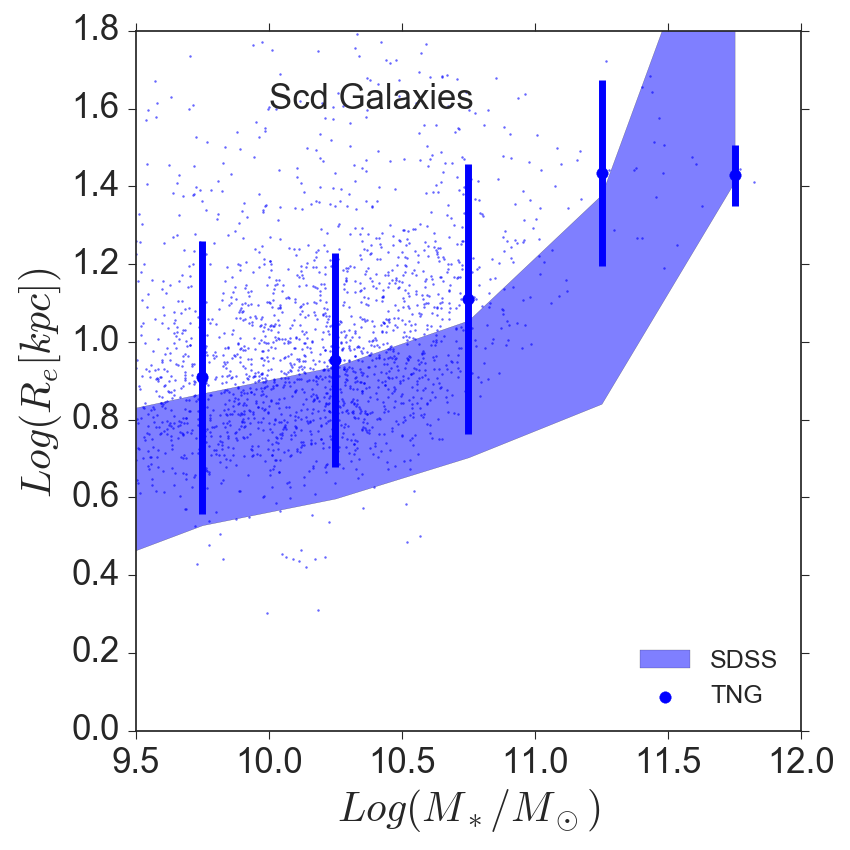}\\
\end{tabular}
\caption{Stellar mass - size relation of observed and simulated galaxies divided in four morphological types as labeled. The shaded regions indicate the observed relations. Small dots indicate individual TNG galaxies and the large dots with error bars are median values and scatter.}
\label{fig:mass_size_morph}
\end{figure*}

In figure~\ref{fig:mass_size_morph} we now explore the mass-size relations divided in finer morphological classes. We also do find a remarkable agreement between observed and simulated galaxies for all the morphological types. The median sizes in TNG generally within the 1-$\sigma$ confidence interval of the observations.  Only simulated late-type spirals (Scd) seem to have a median size at fixed stellar mass 1-$\sigma$ larger than observed. This can be also visually appreciated in figure~\ref{fig:stamps_Scds}. Simulated Scd galaxies tend to have more extended star-forming clumpy disk. Although this was not an issue for the CNN since the features of late-type systems are present, it appears in the scaling relation.

\section{Stellar mass functions of different Hubble types}
\label{sec:SMFs}

\begin{figure*}
\includegraphics[width=0.95\textwidth]{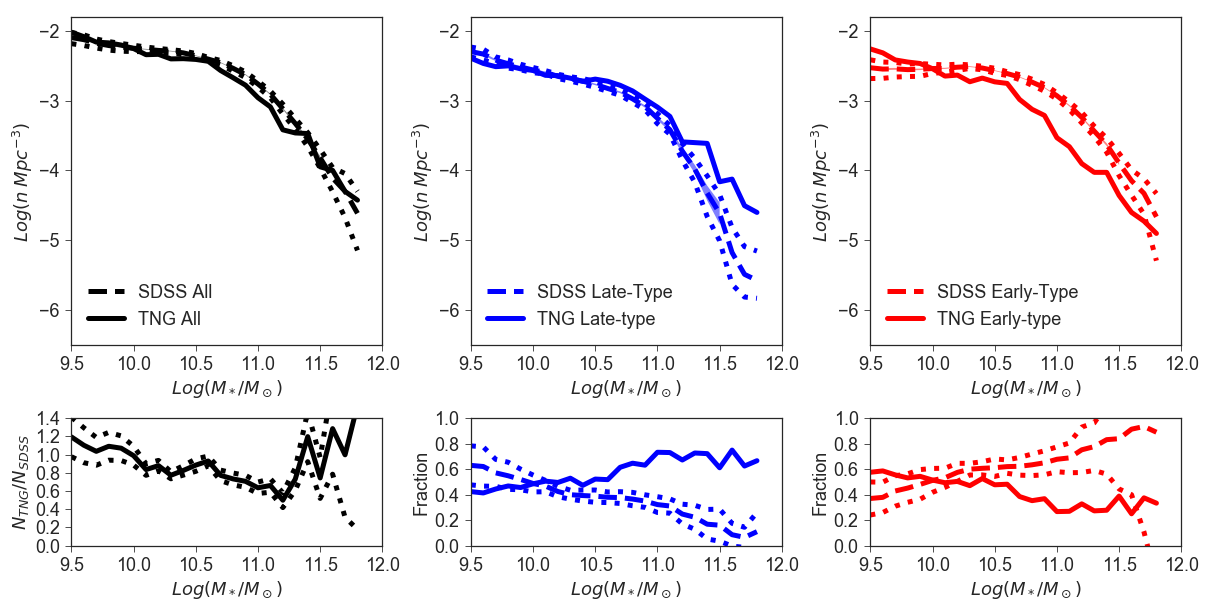}
\caption{Stellar Mass Functions of all (top left panel), early (top right panel) and late-Type (top middle panel) galaxies. The solid lines show the IllustrisTNG measurements and the dashed lines are in SDSS. The dotted lines show the variations due to volume and the shaded regions are Poisson errors. The bottom panels show the fraction of early and late type galaxies as a function of stellar mass in the observations (dashed lines) and in the simulations (solid lines). The dotted lines show the maximum fluctuation in the relative abundances due to volume.}
\label{fig:mass_funct}
\end{figure*}

We explore in this section the stellar mass functions (SMFs) of the different morphologies in the observations and in the simulations. Figure~\ref{fig:mass_funct} shows first the total stellar mass function as well as the SMF of early and late-type galaxies. The bottom panels of figure~\ref{fig:mass_funct} indicate the fraction of early and late-type galaxies as a function of stellar mass as well as the ratio between simulated and observed galaxies in bins of stellar mass. The total stellar mass function is in excellent agreement with the results of~\cite{2017MNRAS.467.2217B}, i.e. to the 20-40 per cent level. This is not too surprising given that the TNG model was designed to improve upon the original Illustris in matching the SMF at z = 0~\citep{2018MNRAS.473.4077P}. 

Surprisingly, when considering the SMF divided by two broad morphological types, we measure opposite trends in the simulated and observational samples. In the SDSS, the high mass end is dominated by early-type galaxies as reported by many previous works (e.g. \citealp{2013MNRAS.436..697B}). However, in TNG the high mass end of the stellar mass function appears to be dominated by disk galaxies. As described in section~\ref{sec:obs}, the volume probed by the simulations is $\sim40$ times smaller than in the observations. The morphological mix at the high mass end could be affected by small statistics. In order to evaluate the impact of this, we recompute the SMF in the SDSS in 40 smaller volumes. The result is shown with dashed lines in figure~\ref{fig:mass_funct}. The difference measured between observations and simulations is significantly larger than the variations caused by measuring abundances in smaller volumes. Therefore, even if TNG is able to produce realistic morphologies, this result suggests that the abundances show important discrepancies with the observations. 

\begin{figure*}
\begin{tabular}{|c|c|}
      \includegraphics[width=0.45\textwidth]{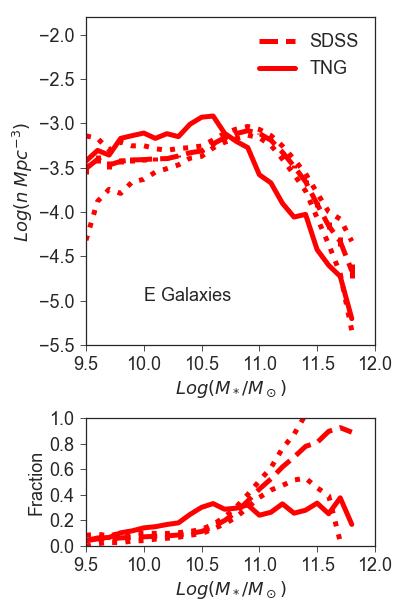} & \includegraphics[width=0.45\textwidth]{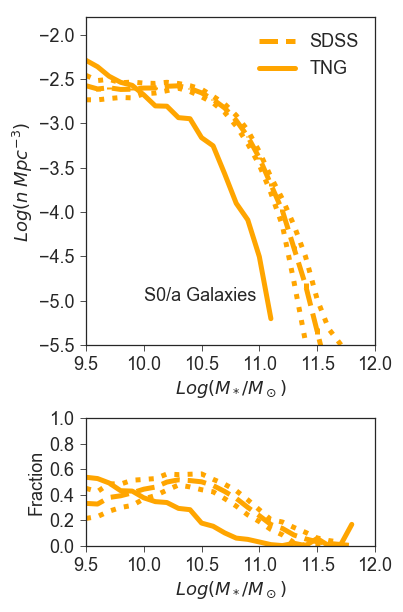}\\
   \includegraphics[width=0.45\textwidth]{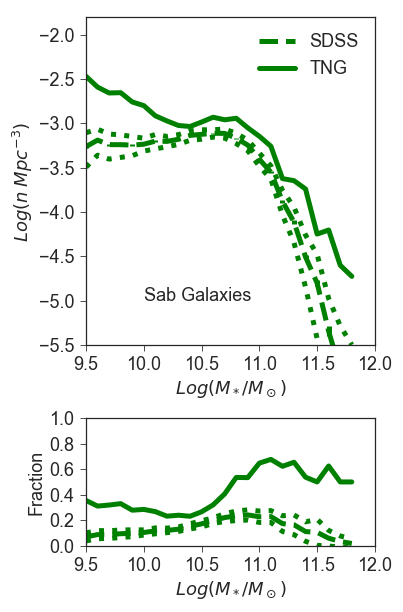} & \includegraphics[width=0.45\textwidth]{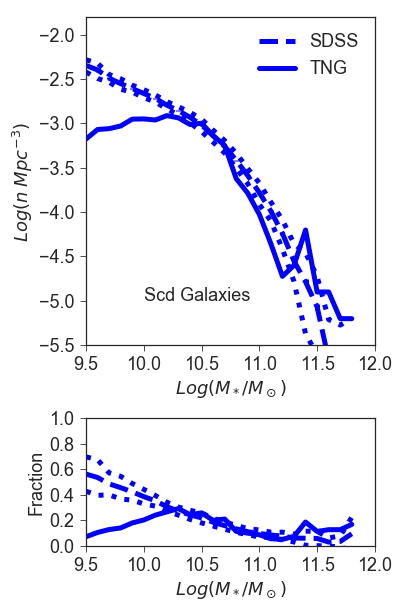}\\
\end{tabular}
\caption{Stellar mass functions of different morphological types as labeled. The solid lines indicate the TNG simulations and the dashed-lines show the measurements in the SDSS. The small panels show the fractions with respect to the total as function of stellar mass. }
\label{fig:mass_funct_morph}
\end{figure*}

We explore further the origins of this discrepancy by dividing the sample in finer morphological classes as described in section~\ref{sec:deep}. The results are shown in figure~\ref{fig:mass_funct_morph}. It confirms that both, the abundances of elliptical and S0/a galaxies are under estimated in the simulations. For a typical mass of $log(M_*/M_\odot)\sim11$, the number densities are a factor of $\sim8$ smaller in TNG for elliptical and S0/a galaxies. There is a comparable excess of early-type spiral galaxies (Sabs). This discrepancy creates the opposite trends reported in figure~\ref{fig:mass_funct} between the early and late-type populations. 

\begin{figure}
\includegraphics[width=0.45\textwidth]{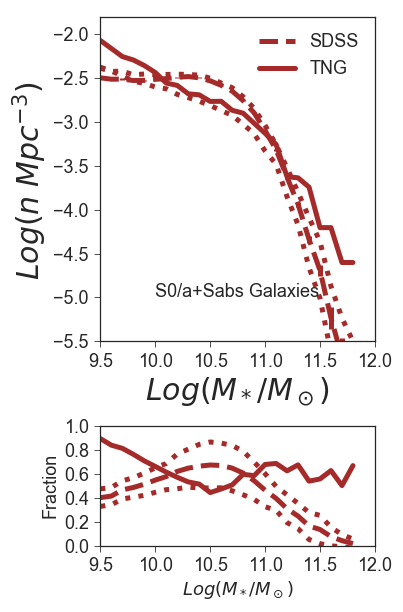}
\caption{Stellar mass function of both S0s and Sab galaxies. The solid lines indicate the TNG simulations and the dashed-lines show the measurements in the SDSS. The small panel show the fractions with respect to the total as function of stellar mass. }
\label{fig:SMF_S0_Sab}
\end{figure}

The difference between S0s and Sa is subtle and separating between the two populations is well known to be challenging. As detailed in the previous sections, the difference between the two classes resides essentially in the size and structure of the disk component. Sa galaxies are expected to have more features in the disk such as spiral arms. This is not always easy to appreciate with limited spatial resolution. For example, using a different classification method based on Support Vector Machines and color information, \cite{2013MNRAS.436..697B}, finds more Sa galaxies at the low mass end in detriment of S0s than our measurements here.  However since the same CNN model was used to classify both the simulations and the observations, there in an internal consistency which allows us to safely argue that the discrepancy is real. Also notice that, as shown in the previous section, the scaling relations of S0 and Sab galaxies are different and well reproduced by the simulations. If this was a problem of classification errors in the simulations we would have measured some deviations in the mass-size relations in figure~\ref{fig:mass_size_morph}.  In order to quantify how much of the measured discrepancy between SDSS and TNG is due to S0 galaxies being classified as Sabs, we plot in figure~\ref{fig:SMF_S0_Sab} the SMFs of the two populations together. Simulations and observations become roughly compatible at intermediate masses. Most of the discrepancy at intermediate masses ($10<Log(M_*/M_\odot)<11$) comes then from S0s being classified as early-type spirals. According to figures~\ref{fig:stamps_S0s} and~\ref{fig:stamps_Sabs}, it seems to be because simulated galaxies at these masses, even if they have a noticeable bulge component, they present an extended disk component with more visible features than what is expected for an S0 galaxy. This pushes the CNN classification to later types. 

\begin{figure}
\includegraphics[width=0.45\textwidth]{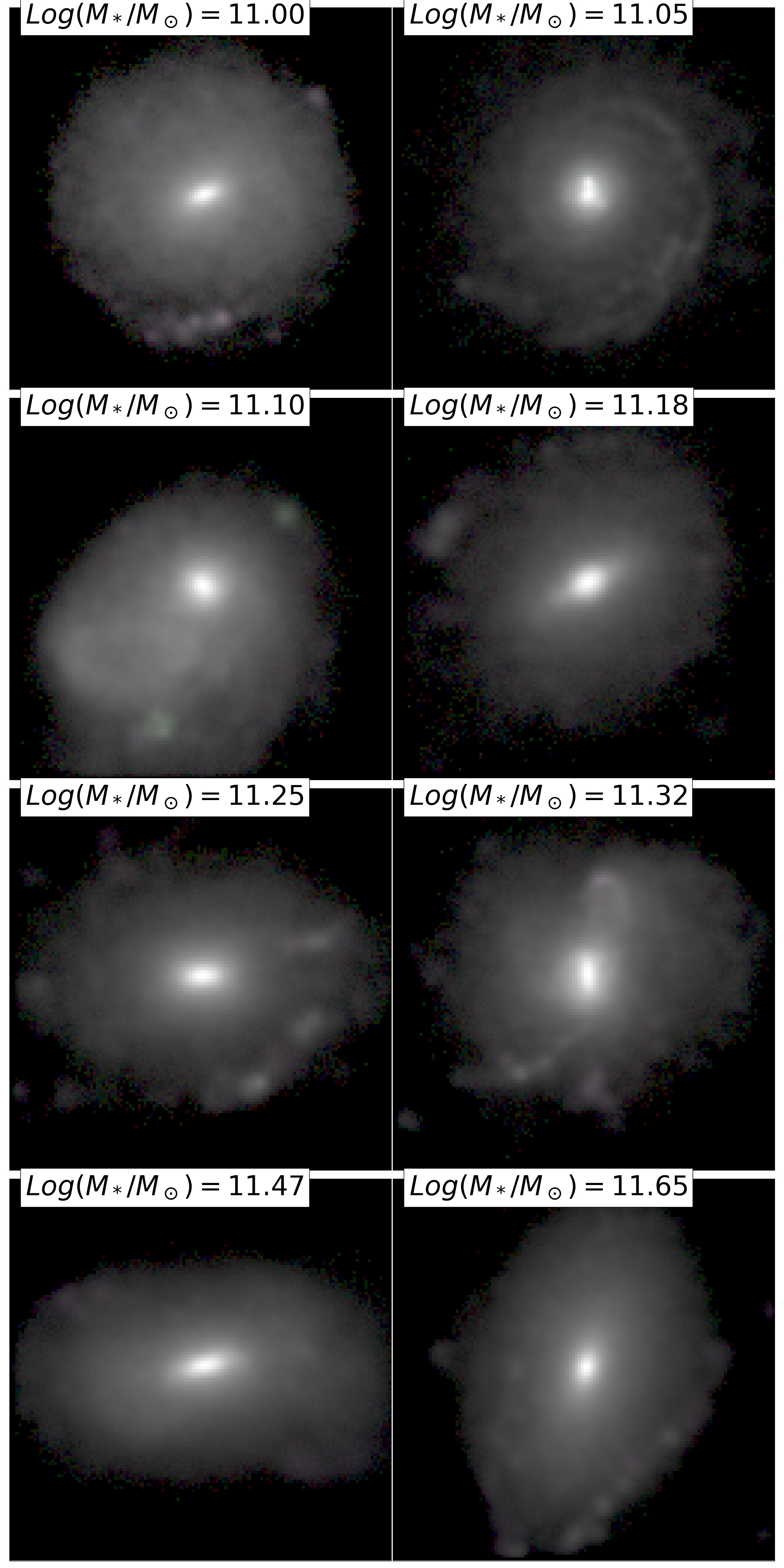}
\caption{Example of stamps of massive galaxies ($M_*/M_\odot>10^{11}$) classified as late-type galaxies by the CNN which dominate the massive end of the SMF in the IllustrisTNG simulation.}
\label{fig:mass_disks}
\end{figure}

Perhaps more interesting is that at the very high mass-end, in which observations are completely dominated by elliptical galaxies, the simulations still present an over-abundance of late-type systems. In figure~\ref{fig:mass_disks}, we show some examples of massive late-type galaxies ($Log(M_*/M_\odot)>11$) in IllustrisTNG. Although the galaxies have a also prominent bulge component, and are on average rounder than typical disks, they also present a clear extended structure which is most probably the feature used by the network to classify the galaxy as a late-type system. These galaxies are in particular different from the typical elliptical galaxies shown in figure~\ref{fig:stamps_Es} and closer to Sab galaxies (figure~\ref{fig:stamps_Sabs}) in the sense that the extended low surface brightness component presents more structure, probably due to on-going star-formation. This probably causes the network to interpret the structure as a disk.

Some additional discrepancies are also seen at the low mass end. The observations show that below $\sim10^{10}$ solar masses, the population is dominated by late-type spirals (Scd). In TNG however early-type spirals (Sabs) with larger bulges also dominate the populations of galaxies. It is known that the spectroscopic SDSS is biased against certain environments due to the issue of fiber placement collisions (see e.g. discussion in~\cite{2018MNRAS.475..624N}). The result is that the number of red/spheroidal satellite galaxies might be underestimated in the observations since they predominantly live in dense regions. It is unlikely though that this effect can explain the measured discrepancy at the low mass end. If the number of missing satellites was significant enough to change the number densities, the global observed and simulated SMFs would not agree as they do at the low mass end (Fig.~\ref{fig:mass_funct}). 

As a final note, one could argue that these discrepancies might be partially caused classification errors in the observations since the training set used lacks faint galaxies (see section~\ref{sec:bnns}). We have checked that this is not the case by computing the abundances of the different morphological types only in the N10 sample. We measure very similar trends as in the whole M15 sample although with more noise given the incompleteness and low statistics. Another potential source of discrepancy could arise from the way images of simulated TNG galaxies are generated. Discreteness effects from the finite particle resolution and smoothing procedure applied to the stars (e.g. as discussed/explored in \citealp{2015MNRAS.447.2753T}) influence the presence of feature/structure that can be interpreted by the CNNs as disks. In future work, we plan to quantify the impact of this by experimenting with the smoothing prescription in low stellar density regimes.

\begin{figure*}
\begin{tabular}{|c|c|}
      \includegraphics[width=0.45\textwidth]{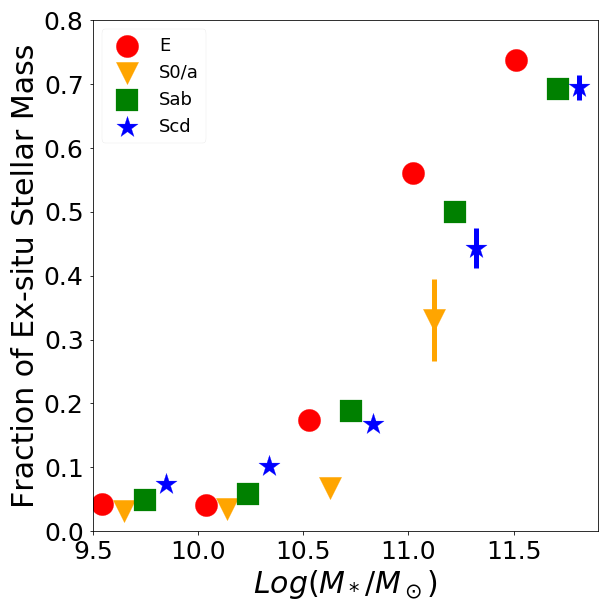} & \includegraphics[width=0.45\textwidth]{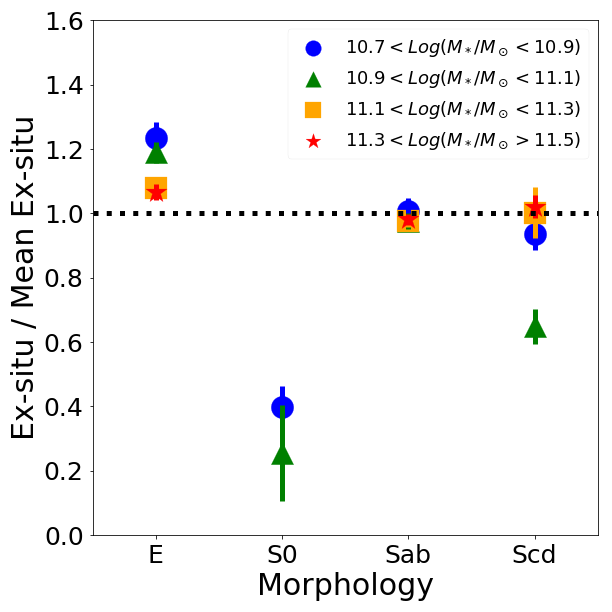}\\
 \end{tabular}
\caption{Fraction of the stellar mass formed outside the galaxies as a function of stellar mass and morphology. The left panel shows the fraction as function of stellar mass at fixed morphological type. The different symbols indicate different stellar mass bins as labeled. The right panel shows the ex-situ mass for a fixed morphological type divided by the mean ex-stiu mass in a given stellar mass bin. The dotted lines indicates a relative fraction of 1 for reference. Error bars are errors on the median values. Only points with more than 10 galaxies are plotted.}
\label{fig:ass_histories}
\end{figure*}

\section{Discussion: the assembly histories of different morphologies}
\label{sec:disc}
The previous sections have shown that the visual morphologies of galaxies in the TNG simulation reproduce fairly well the observed morphologies both in terms of global, visual morphology and scaling relations. However, there are still some discrepancies in the abundances of early and late-type galaxies. It suggests that while the assembly channels of the different morphological types produce a realistic morphological distribution, the relative importance of the different mechanisms does not seem to be correct so the number densities are not well reproduced.

In order to better understand the origin of the morphological classes, as well as possibly of these discrepancies especially at the high mass end, we explore in figure~\ref{fig:ass_histories} some tracers of the assembly histories. In particular, we first look at the contribution of mergers in the stellar mass assembly. The left panel of figure~\ref{fig:ass_histories} shows the fraction of ex-situ stellar mass as a function of stellar mass (Ex-Situ Stellar Mass). By ex-situ we mean stars that have formed not in-situ i.e. from gas condensing within the innermost regions of the observed galaxy (or its main progenitors) but in other galaxies that have been accreted, stripped and that have possibly merged with a galaxy prior to the time of observation (see~\citealp{2016MNRAS.458.2371R} and~\citealp{2018MNRAS.475..648P} for operational definitions and basic results from Illustris and the IllustrisTNG simulations). The ex-situ stellar mass fraction should be a proxy of the importance of mergers in the assembly histories. 

Firstly, as previously shown (e.g. \citealp{2016MNRAS.458.2371R}, ~\citealp{2018MNRAS.475..648P} and reference therein), the ex-situ fraction is a very strong function of galaxy mass. Below$~\sim{10^{10.5}}$ solar masses, the amount of accreted stellar mass is negligible (<15-20\%) for all morphologies. Figure\ref{fig:mass_funct_morph} shows that the majority of low-mass galaxies in TNG have a bulge component (S0 and Sab galaxies account for $\sim80$\% of the number densities). Depending on the bulge-to-total mass fraction, it could be that bulges in these systems might have grown through (also or exclusively) internal processes. Above $\sim10^{10.5}$ solar masses, the amount of accreted stellar mass starts to be significant and reaches almost 80\% at $10^{11.5}$ solar masses, with large galaxy-to-galaxy variations. The scatter in ex-situ fraction at fixed galaxy mass for different morphological types is also large. Yet, we do observe that massive ellipticals have larger ex-situ fractions than Scd, at least around the $10^{11}$ Msun scale: e.g. ~65\% vs. ~45\% at $10^{11.25}M_\odot$. This finding is consistent with the results of~\cite{2017MNRAS.467.3083R}. However, S0 galaxies seem to exhibit lower ex-situ fraction than all other types at all masses. Although this might be a consequence of low statistics, it appears to be a systematic trend at all masses.

We expand on these trends in the right panel. Ex-Situ fractions are plotted at fixed stellar mass for different morphologies, by focusing on the mass regime where the ex-situ contribution is non negligible. The panel shows the relative excess of ex-situ mas fraction of a given morphological type as compared to the average ex-situ mass in a given stellar mass bin. The curves show a weak trend of average ex-situ fraction with morphology in bins of stellar mass, with a strong under abundance of ex-situ mass for massive S0 galaxies (again possibly due to low statistics, as revealed by the SMFs of section~\ref{sec:SMFs}). At the highest mass end, ($>10^{11.3}$), the differences in ex-situ mass fractions across morphological types is very small (curves are essentially flat). This result does not mean that massive elliptical galaxies are not formed through mergers. Indeed the high mass end of the SMF in TNG is populated by a significant fraction of ellipticals which are likely formed through mergers given the large fraction of ex-situ mass. However, it looks like a significant fraction of mergers does not produce early-type galaxies but the final morphology is an Sab galaxy which therefore dominate the high mass end of the stellar mass function. The figures show that there are no strong obvious differences in the assembly histories of different morphologies, suggesting that subtle differences in assembly histories may be responsible for changes in the morphological type. If the merger history is similar, what determines that a massive galaxy will end up as Sab or Elliptical? The answer might be in the properties of the mergers and accretion events. 

\begin{figure*}
\begin{tabular}{|c|c|}
      \includegraphics[width=0.45\textwidth]{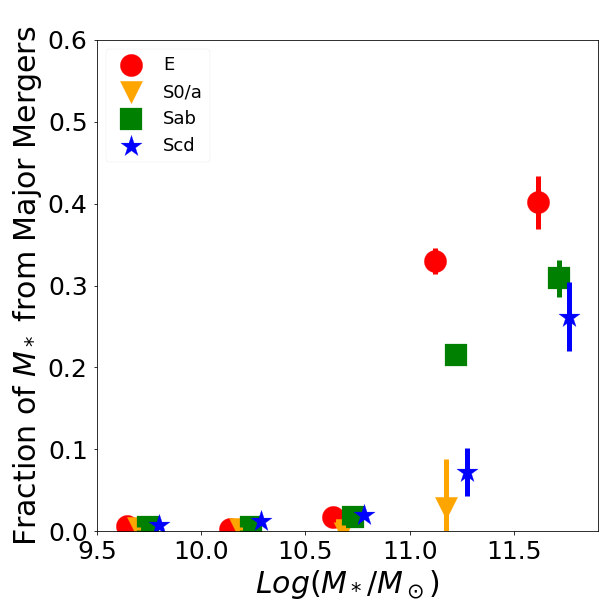} & \includegraphics[width=0.45\textwidth]{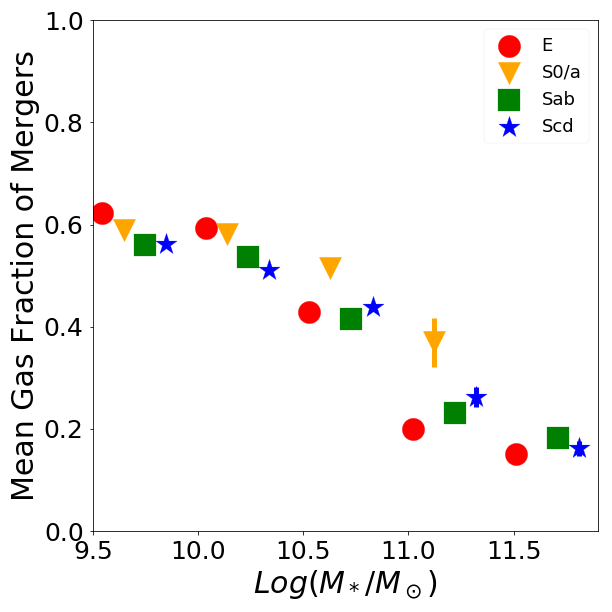}\\
 \end{tabular}
\caption{Left panel: Fraction of the stellar mass formed through major mergers as a function of stellar mass and morphology. Right panel: Mean gas fraction integrated over all the merger events as a function of morphology and stellar mass. Error bars are errors on the median values. Only points with more than 10 galaxies are plotted.}
\label{fig:merg_props}
\end{figure*}

In figure~\ref{fig:merg_props}  we plot the fraction of mass coming from major mergers only (left panel) as well as the median gas fraction involved in the mergers (right panel). Consistently with the the ex-situ mass fraction findings, below $10^{10.5}M_\odot$ galaxies mostly accrete stars via minor mergers. The relative contributions of major merges is again a strong function of galaxy mass above $\sim10^{11}$ solar masses. At the high mass end, the figure suggests that the merger mass ratio might be a relevant factor in determining the final morphology in the simulation. Elliptical galaxies tend to have a larger fraction of stellar mass coming from major mergers than early-type spirals (0.4 vs. 0.2). Surprisingly there is little difference in terms of gas fraction, although elliptical galaxies tend to be formed in slightly dryer mergers, whereas S0 in more gas-rich events. 

The tentative picture that seems to emerge is that at the low mass end ($<10^{10.5}M_\odot$) the assembly history has very little to null impact in setting galaxy morphologies. At the massive end ($>10^{10.5}M_\odot$), if the accretion and merger histories contribute to determine galaxy morphologies, their manifestations are subtle. The amount and types of mergers do manifest differently for different morphological types, but this is the case only for those galaxy masses where there is significant ex-situ contribution and with relatively weak trends. Furthermore, in our model, a larger fraction of major mergers will tend to form an elliptical galaxy. In the majority of the cases however (recall that the massive end of the SMF is dominated by Sab galaxies in TNG) the galaxies will end up with a disky morphology even if $\sim$50\% of their stellar mass is accreted. It is unclear if this is because the mergers are not big enough to destroy the disk or because there is still a fairly large amount of available gas that is re-accreted. In fact, it is likely that morphological transformations are also associated to the nature and strength of the feedback mechanisms in place, particularly the feedback from the central super massive black holes, which may act in conjunction with galaxy mergers to set galaxy morphologies. We postpone to future work the investigation of the relation between morphological types and feedback history. 

A possible resulting effect of what seen thus far seems to be that the TNG simulations lack of an efficient way to form lenticular galaxies. It seems that either the disk is fully destroyed or it remains too important and featured to be considered an S0. A possible explanation could be that S0s are preferentially formed in high density environments such as clusters. Some observational works have shown that there is a larger fraction of S0 galaxies in clusters as compared to the field  (e.g.~\citealp{2009ApJ...690...42M, 2013MNRAS.428.1715H}).. Since the TNG volume is relatively small  ($\sim100$ $Mpc^{3}$), the number of halos at the cluster scale is small (about 10 haloes more massive than $10^{14}M_\odot$  in total mass) and so the environmental effects may not be well represented in comparison to the Universe demographics. This could be investigated with the larger TNG300 volume, however at the expenses of resolution. Here we point out that a quick inspection of the fraction of different morphological types as a function of halo mass in the SDSS sample shows that S0 and Sab galaxies live in very similar environments. Elliptical galaxies do tend to live in denser environments but their number densities are better reproduced. It is therefore unlikely that the discrepancy we measure can be fully explained by environmental considerations.

 Finally, the low mass end of the SMF also presents some discrepancies. The number densities of early-type spirals are too high in the simulations in comparison to the observations. Moreover, there is an over abundance of low-mass elliptical galaxies whereas the SDSS SMF is completely dominated by Scd galaxies with almost no bulge. This suggests that bulges are formed too efficiently in low mass galaxies in the TNG100 simulation and such discrepancy calls for future analyses with simulations at better resolution, e.g. TNG50.

\section{Summary and conclusions}

We have quantitatively analyzed the visual morphologies of galaxies at $z\sim0$ in the IllustrisTNG simulation. We have trained a CNN on detailed visual morphologies estimated on $14,000$ galaxies in the SDSS and applied the same network to classify a complete sample of $12,000$ galaxies in TNG with stellar mass larger than $10^{9.5}$ solar masses. In order to produce images with similar properties than in the observations, the output of the simulations was post-processed with a radiative transfer code to create mock observations with realistic instrumental effects. Our morphological classes include early-type galaxies, in turn divided in ellipticals E and lenticulars S0/a, and late-type galaxies, in turn classified as early-type spirals (Sab) and late-type spirals (Scd), the latter including irregulars.

Our main results are:
\begin{itemize}
\item The TNG simulation reproduces well the diversity of morphologies in the local universe. A CNN trained on the SDSS is able to find galaxies in different morphological types in the simulation with high confidence for $\sim95\%$ of the sample. Even if some differences might exist, it means that the main features learned by the networks at the SDSS resolution are present both in the simulations and in the observations. An analysis of these features shows indeed that they cluster similarly. 
\item The mass-size relations of simulated galaxies reproduce well the slope and the normalization of the observed relations for all morphological types. This includes the global trends for the early and late type populations, but also when galaxies are divided in finer morphological classes. The scatter at fixed stellar mass remains slightly larger in the simulations. This is a significant improvement as compared with the original Illustris run. The only exception to this are late-type spirals (Scd), which are ~0.2dex larger in TNG than in SDSS    
\item We measure some discrepancies in the stellar mass functions divided by morphological type both at the low and high mass end. The high-mass end of the SMF in the simulation is dominated by late-type systems while early-type galaxies are more abundant in the SDSS. We show that this is due to a lack of lenticular galaxies in the TNG simulation at intermediate masses and ellipticals at the high mass end, probably because there is still too much available gas to build disks. At the low mass end, early-type spirals dominate in the simulation while in the observations most of the galaxies are late-type spirals (with no bulge). It suggests that bulges are formed too efficiently in low mass galaxies in the simulation. Remaining differences in our synthetic images vs. SDSS imaging, namely the exclusion of nearby companions and the treatment of stellar smoothing, could influence these points and can be directly assessed in future work.
\item At the low-mass end ($<10^{10.5}M_\odot$) the merger histories of galaxies have no manifest impact in setting morphologies There is a small  (but statistically significant) difference between the merger histories of massive galaxies with different morphologies. At $log(M_*/M_\odot)\sim11$, the fraction of accreted mass through mergers in elliptical galaxies is $\sim10\%$ larger than in spiral galaxies. The nature of the mergers is also different. Elliptical galaxies have on average $\sim20\%$ more stellar mass coming from major mergers. However, the influence of assembly history in setting the other morphological types remains unclear.
\end{itemize}

This work has shown that there is potentially interesting information to be learned by consistently comparing simulations and observations in the same observational frame. In particular, detailed morphologies which can be now estimated with high accuracy can provide additional constraints on the physical processes driving galaxy assembly  and help improving the next generation of simulations. In future work we will extend this analysis to high redshift using the high resolution TNG50 simulations~\citep{2019arXiv190205554N, 2019arXiv190205553P} and Hubble Space Telescope imaging. We also plan to explore Generative Models as a more general way to confront models of galaxy formation and observations. 

\label{lastpage}

\end{document}